\documentclass[11pt]{article}
\usepackage{amssymb,latexsym,amsmath,graphicx}
\usepackage{subfig}
\usepackage{latexsym}
\usepackage[mathscr]{euscript}
\usepackage{soul}
\usepackage{float}
\usepackage{tabularx,ragged2e,booktabs,caption}
\usepackage{multirow}
\renewcommand{\textwidth}{6.5in}

\newcommand{\argmin}{\operatornamewithlimits{argmin}}

\newtheorem{thm}{Theorem}
\newtheorem{lem}{Lemma}
\newtheorem{rem}{Remark}
\newtheorem{cor}{Corollary}

\usepackage{epstopdf}
\AppendGraphicsExtensions{.tiff}

\begin{document}
	\begin{center} 
		\Large{{\bf Statistical classification for partially observed functional data via filtering}} 
	\end{center}
	
	% \vskip .1 in
	
	\begin{center}
		%By \\
		Majid Mojirsheibani\footnote{Corresponding author. ~Email:  majid.mojirsheibani@csun.edu
			
			This work is supported by the NSF Grant DMS-1407400 of Majid Mojirsheibani.},
		My-Nhi Nguyen\footnote{Email: mynhi.nguyen.324@my.csun.edu},
		and Crystal Shaw\footnote{Email: c.shaw@ucla.edu}
		
		Department of Mathematics, California State University Northridge, CA, 91330, USA$^{1,2}$\\
		Department of Biostatistics, University of California Los Angeles, CA, 90095, USA$^{3}$
	\end{center}
	
	\begin{abstract}
		This article deals with the problem of functional classification for $L^2$-valued random covariates when some of the covariates may have missing or unobservable fragments. Here, it is allowed for both the training sample as well as the new unclassified observation to have missing fragments in their functional covariates. Furthermore, unlike most previous results in the literature, where covariate fragments are typically assumed to be missing completely at random, we do not impose any such assumptions here.  Given the observed segments of the curves, we construct a kernel-type classifier which is quite straightforward to implement in practice. The proposed classifier is constructed based on $d$-dimensional covariate vectors, obtained from the original covariate curves (by moving from $L^2$ to the space $\ell_2$), where $d$ itself is a parameter that has to be estimated. To estimate various parameters, we employ a random data-splitting approach which is easy to implement. We also establish the strong consistency of the proposed classifier and provide some numerical examples to assess its performance in finite sample problems.
	\end{abstract}
	
	 \allowdisplaybreaks	
\section{Introduction}
The problem of statistical classification and pattern recognition with functional covariates has received considerable  attention in recent years. This is particularly true when the data are fully observable. In a standard two-group classification problem, this amounts to considering the random pair $(\boldsymbol{\chi} , Y)$, where $\boldsymbol{\chi}$ is a functional covariate taking values in some metric space $({\cal M}, d)$ and $Y\in \{0,1\}$, called the class membership or class variable,  has to be predicted based on $\boldsymbol{\chi}$.  
Here, one would like to find a classifier (a function) $g:{\cal M}\to \{0,1\}$  for which the misclassification error,  $L(g) :=P\{g(\boldsymbol{\chi})\neq Y\}$, is as small as possible. The optimal classifier, i.e., the classifier with the lowest misclassification error, is given by $g_{\mbox{\tiny B}}(\chi)=
1$ if $P\{Y=1|\boldsymbol{\chi}=\chi\}>1/2$, and $g_{\mbox{\tiny B}}(\chi)=
0$ otherwise; see, for example,  C\'erou and Guyader (2006), Abraham et al. (2006), as well as the monograph by Devroye, et al. (1996; Ch. 2).
%In passing we also note that if $\mu$ represents the probability measure of $\boldsymbol{\chi}$, i.e., $\mu(B)=\mathbb{P}\{\boldsymbol{\chi}\in B\}$ for each Borel set $B\subset {\cal M}$, then the joint probability distribution of $(\boldsymbol{\chi},Y)$ is completely determined by $\mu$ and the regression function $\eta(\boldsymbol{\chi}) :=\mathbb{P}\{Y=1 | \boldsymbol{\chi}=\chi\}=\mathbb{E}[Y | \boldsymbol{\chi}=\chi]$.
Although we have presented our setup for the popular binary case where $Y\in\{0,1\}$, our discussions and results in this paper can be generalized in a straightforward manner to the multi-group classification problem where $Y \in \{1, 2, \dots, C\},$ for some positive integer $C\geq 2$. \\

\noindent
In practice the optimal classifier $g_{\mbox{\tiny B}}$ is virtually always unknown (because the conditional probability $\mathbb{P}\{Y=1|\boldsymbol{\chi}=\chi\}$ is not available) and one only has access to a set of $n$ independent and identically distributed (iid)  data values $\mathbb{D}_n = \{(\boldsymbol{\chi}_1 , Y_1),\dots,(\boldsymbol{\chi}_n , Y_n)\}$ from the underlying distribution of $(\boldsymbol{\chi} , Y)$. 
The task of classification is then to use the data $\mathbb{D}_n$ to construct a classification  rule $g_n$ that can predict the class membership, $Y$, of a new curve $\boldsymbol{\chi}$ with low error rates. A variety of techniques have been proposed for the classification of functional data in the literature. One may divide these techniques into roughly two types: (a) those approaches that use the whole curve $\boldsymbol{\chi}$ to predict $Y$ and (b) those that use the {\it filtered} curves to carry out classification; here, a filtered curve is a representation of a curve in the form of a vector. Relevant results corresponding to the approach used under (a) include the nonparametric functional approach of Ferraty and Vieu (2003), the nearest neighbor method used by C\'erou and Guyader (2006), the kernel classifier of Abraham et al. (2006), the depth-based classifier of L\'opez-Pintado and Romo (2006), the robust functional classification of Cuevas et al. (2007), 
the wavelet approach of Chang et al. (2014), the robust functional classification of Alonso et al.(2014),
and the work of Meister (2016) on the optimality properties of kernel regression and classification with functional covariates taking values in a general complete separable metric space.  \\

\noindent
On the other hand, relevant work under (b)  includes the discrimination method of Hall et al. (2001), the functional classification method of Biau et al. (2005),  the results of Leng and M{\"u}ller (2006) on the classification of gene expression data as well as that of Song et al. (2008), the wavelet approach of Berlinet, et al. (2008), the componentwise classification approach of Delaigle, et al. (2012), the classification method in Delaigle and Hall (2012), the {\it depth-depth} plot approach of Mosler and Mozharovskyi (2017), and
the functional classification method of Dai and M{\"u}ller (2017). Some other relevant results (but in the context of functional regression) include the work of Cai and Hall (2006) on prediction in functional linear regression, the results of Hall and Horowitz (2007) on the estimation of a slope function in functional linear regression, and those of Yao and M{\"u}ller (2010) on functional quadratic regression.  \\

\noindent
In this paper we employ methods that primarily fall under (b) above. More specifically, assuming that the functional covariates take values in a separable Hilbert space (and using the fact that such spaces are isomorphic to the space $\ell_2$), the functional covariates will be replaced by $d$-dim vectors where $d\equiv d(n)$ is to be determined by the data; here, $d(n)\to\infty$, as $n\to\infty$. For the missing data framework,  we follow the setup proposed by Bugni (2012), which has also been employed by  Kraus (2015) as well as  Mojirsheibani and Shaw (2018); this is described in Sections \ref{back}. In section \ref{MAIN} we propose a kernel classifier, under multiple missing patterns, and study its asymptotic properties.  Some numerical examples are also given; these appear in Section \ref{Example}. All proofs are deferred to Section \ref{proof}.

%A sample-based classification rule $g_n$ is said to be consistent if its error,  $L(g_n) :=\mathbb{P}\{g_n(\boldsymbol{\chi})\neq Y | \mathbb{D}_n\}$, is such that $\mathbb{E}[L(g_n)]\to L(g_{\mbox{\tiny B}})$, as $n\to\infty$. On the other hand, if $L(g_n)\to L(g_{\mbox{\tiny B}})$ with probability one, then $g_n$ is said to be strongly consistent.\\

\section{Partially observed curves and the setup}
\subsection{Background}\label{back}
In standard functional classification, one typically assumes that each observation (covariate) $\boldsymbol{\chi}(t)$ is a smooth curve on some compact domain $\mathcal{I}\subset \mathbb{R}$. Furthermore, the great majority of existing results assume that $\boldsymbol{\chi}(t)$ as well as $\boldsymbol{\chi}_i(t)$, $i=1,\dots,n$, do not have any missing or unobservable fragments over the domain $\mathcal{I}$. In contrast, here we allow $\boldsymbol{\chi}$ to be possibly missing (unobservable) on some subset(s) of its domain, i.e., the situation where one may only be able to observe certain segments of the full curve $\boldsymbol{\chi}$. In fact, to the best of our knowledge, the problem of functional classification with partially observed covariates has received very little attention in the literature.  
Some key results along these lines in the literature appear to include the work of Delaigle and Hall (2013) who consider a quadratic discriminant classifier for censored functional data based on 
%the spectral decomposition of an empirical covariance function obtained from 
the observed fragments of covariates with overlapping domains that are not {\it too short}. Another relevant result here is that of Kraus (2015) who proposes methods to estimate
parameters and to carry out principal component analysis with missing data. These authors assume that the missingness is independent of the covariate and response variables, which amounts to having covariates missing completely at random (MCAR). In this paper we do not impose any MCAR assumptions.
More specifically, let $(\Omega, {\cal A},\mathbb{P})$ be the underlying probability space and let $\mathcal{M}$ be the space of square-integrable functions $L^2(\mathcal{I})$, where ${\cal I}$ is an interval on the real line. Therefore, $\boldsymbol{\chi}$ is a random function on  $(\Omega, {\cal A},\mathbb{P})$ with values (i.e., with sample paths) in $L^2({\cal I})$. But, instead of observing the full curve $\boldsymbol{\chi}:\Omega\to L^2({\cal I})$, one might only be able to observe certain segments of the curve denoted by $\boldsymbol{\chi}|_{s}$, i.e., the restriction of the curve $\boldsymbol{\chi}(t)$ to $t\in s\subset \mathcal{I}$. \\

\noindent
To set up our framework for possible missing patterns in the curve $\boldsymbol{\chi}$, we follow the setup proposed by Bugni (2012). This method is also employed by  Kraus (2015) who considers principal component analysis with missing data. 
In Bugni's (2012) setup, it is assumed that for a fine enough partition of ${\cal I}$ into $J<\infty$ subintervals ${\cal I}_1,\dots, {\cal I}_J$, each sample function of $\boldsymbol{\chi}$ is either completely observed or completely unobserved within each of these $J$ subintervals.  Some examples of such functional variables can be found in  Bugni (2012). In the rest of this paper we assume that there are $M< 2^J$ possible missing patterns in the data where $M$ is usually much smaller than $2^J$. Therefore, under the $k$-th pattern, one observes the fragment $\boldsymbol{\chi}|_{s_k}$, $k=1,\dots,M$. Next, let $\delta$ be the $\{1,\dots, M\}$-valued random variable defined as
\[
\delta = k ~~~\mbox{if pattern $k$ (i.e., the fragment $\boldsymbol{\chi}|_{s_k}$) is observed},~k=1,\dots,M.
\]
Therefore, if we let $\boldsymbol{\chi}^{(\delta)}$ represent the observed covariate fragment, then it can be written as $\boldsymbol{\chi}^{(\delta)}=\sum_{k=1}^M  \boldsymbol{\chi}|_{s_k}\,I\{\delta=k\}$, where, without loss of generality, one may take $s_1=\mathcal{I}$, i.e., the case where the entire curve $\boldsymbol{\chi}$ is observable on $\mathcal{I}$. In passing we also note that when pattern $k$ is observed, then a classifier is any function of the form~ $g_k: L^2(s_k) \to \{0,1\}$. Therefore, given $M$ possible missing patterns, any classifier is necessarily of the form
\begin{equation}\label{E2}
\Gamma_0(\boldsymbol{\chi}^{(\delta)})= \sum_{k=1}^M I\{\delta=k\}\cdot g_k(\boldsymbol{\chi}|_{s_k}),~~\mbox{for some  $g_k: L^2(s_k) \to \{0,1\},~k=1,\dots,M$.}
\end{equation}
As for the theoretically best classifier for the current setup (with missing fragments in $\boldsymbol{\chi}$), let
%\begin{equation}\label{E3}
$\phi_k(\chi|_{s_k})= E\left[(2Y-1) I\{\delta=k\}\,\big|\, \boldsymbol{\chi}(t)=\chi(t),~ t\in s_k\right]$
%\end{equation}
and consider the classifier 
\begin{equation}\label{E4}
\Gamma_0^{\mbox{\tiny B}}(\boldsymbol{\chi}^{(\delta)}) = \sum_{k=1}^M I\{\delta=k\}\cdot I\{\phi_k(\boldsymbol{\chi}|_{s_k})>0\}.
\end{equation}
The following result shows that $\Gamma_{0}^{\mbox{\tiny B}}$ is the optimal classifier (it has the lowest error).
\begin{thm}\label{A} {\it [Mojirsheibani and Shaw (2018; Theorem 1).]}\\
	The classifier $\Gamma_{0}^{\mbox{\tiny B}}$ defined by (\ref{E4}) is optimal in the sense that for any other classifier $\Gamma_0$, one has
	$\mathbb{P}\{\Gamma_0^{\mbox{\tiny B}}(\boldsymbol{\chi}^{(\delta)})\neq Y\} \leq \mathbb{P}\{\Gamma_0(\boldsymbol{\chi}^{(\delta)})\neq Y\}$.
\end{thm}
We note that since any classifier $\Gamma$ is of the form (\ref{E2}), Theorem \ref{A} implies that 
\begin{eqnarray*}
\mathbb{P}\left\{\Gamma_0^{\mbox{\tiny B}}(\boldsymbol{\chi}^{(\delta)})\neq Y\right\} 
=
 \inf_{g_k:~ L^2(s_k)\to \{0,1\},~k=1,\dots,M~}\,
\mathbb{P}\left\{\mbox{$\sum_{k=1}^M$} I\{\delta=k\}\cdot g_k(\boldsymbol{\chi}|_{s_k})\neq Y\right\}.
\end{eqnarray*}

Since $\boldsymbol{\chi} \in L^2(\mathcal{I})$, which is a separable Hilbert space, it can be expressed by the expansion $\boldsymbol{\chi}(t) = \sum_{j=1}^{\infty} X_j\psi_j(t)$, where $\lbrace\psi_1, \psi_2,...\rbrace$ is a complete orthonormal basis for $L^2(\mathcal{I})$ and $X_j = \langle\boldsymbol{\chi},\psi_j\rangle:=\int_{\mathcal{I}} \boldsymbol{\chi}(t)\psi_j(t)dt$. Here the infinite sum converges in $L^2$. 
Similarly, given the data $(\boldsymbol{\chi}_i, Y_i),~ i = 1,...,n$, we can write $\boldsymbol{\chi}_i(t) = \sum_{j=1}^{\infty} X_{ij}\psi_j(t)$, with $X_{ij} = \int_{\mathcal{I}} \boldsymbol{\chi}_i(t)\psi_j(t)dt$. Since any infinite-dimensional separable Hilbert space is isomorphic to the space $\ell_2=\big\lbrace{\bf x}= (x_1,x_2,\dots)\big\vert \sum_{i=1}^{\infty} |x_i|^2 < \infty\big\rbrace$, the \textit{scores} $X_{ij},~ j\geq1$, are used as surrogates for the datum $\boldsymbol{\chi}_i$ in the literature in the sense that knowing $\bold{X}_i:=(X_{i1},X_{i2},\dots)$ is the same as knowing $\boldsymbol{\chi}_i$; see, for example, Hall et al (2001) or Biau et al (2005). This fact is also formalized in part (ii) of Theorem \ref{B1} of the current paper for the particular case of classification with missing functional covariates.\\

To simplify our presentation, we first look at the oversimplified case where there is only one missing pattern. More specifically, write $\mathcal{I}=[a,b]= [a,c]\cup(c,b]$, for some $a<c<b$, where $\boldsymbol{\chi}(t)$ may be missing on $(c,b]$ only. Therefore, we have the expansions
\begin{eqnarray}
{\boldsymbol{\chi}(t)}
&=&
\sum_{j=1}^{\infty} \left\langle\boldsymbol{\chi},\psi_j\right\rangle_{[a,b]}\psi_j(t)
%\nonumber\\
%&=&
\,=\,  \sum_{j=1}^{\infty}\left[\int_a^c \boldsymbol{\chi}(t)\psi_j(t)dt
+ \int_c^b \boldsymbol{\chi}(t)\psi_j(t)dt\right]\psi_j(t) \nonumber
\\
&=&
\sum_{j=1}^{\infty}\big(\left\langle\boldsymbol{\chi},\psi_j\right
\rangle_{[a,c]}
+ \left\langle\boldsymbol{\chi},\psi_j\right\rangle_{[c,b]}\big)\,\psi_j(t).\nonumber 
%\label{E5} 
\end{eqnarray}
Now the surrogate vector of score functions can be written as
%As in the typed pages
\begin{eqnarray*}
	{\bf X}&=&(X_1, X_2, \dots) := \big(\langle \boldsymbol{\chi}, \psi_1\rangle_{\mbox{\tiny $[a,b]$}}\,,\, \langle \boldsymbol{\chi}, \psi_2\rangle_{\mbox{\tiny $[a,b]$}}\,, \dots\big)\\
	&=& \big(\langle \boldsymbol{\chi}, \psi_1\rangle_{\mbox{\tiny $[a,c]$}}\,,\, \langle \boldsymbol{\chi}, \psi_2\rangle_{\mbox{\tiny $[a,c]$}}\,, \dots\big)
	+ \big(\langle \boldsymbol{\chi}, \psi_1\rangle_{\mbox{\tiny $[c,b]$}}\,,\, \langle \boldsymbol{\chi}, \psi_2\rangle_{\mbox{\tiny $[c,b]$}}\,, \dots\big)\\
	&=:&(Z_1,Z_2, \dots) + (V_1,V_2, \dots)  \\
	&=:& {\bf Z}+{\bf V}, 
\end{eqnarray*}
where $\textbf{V}$ may be missing, but not $\textbf{Z}$. Here, we note that if $\textbf{V}$ is not missing then ${\bf X} = \textbf{Z}+\textbf{V}$ is fully observable, otherwise the classification will be based on $\textbf{Z}$ only. In fact, if we put
%Page 7
\begin{equation*}
\phi_1({\bf X}) =\mathbb{E} \left [ (2Y-1)I\{\delta=1\}\big\vert{\bf X}\right]~~
\mbox{and}~~
\phi_0({\bf Z}) =\mathbb{E} \left [ (2Y-1)I\{\delta=2\}\big\vert{\bf Z}\right],
\end{equation*}
where $\delta=1$ if ${\bf X}$ is fully observable (otherwise $\delta=2$), and define the classifier
\begin{equation*}
\Gamma^{\mbox{\tiny B}}({\bf X}^{(\delta)})=I\{\delta=1\} I\big\{\phi_1({\bf X})>0\big\}
+I\{\delta=2\}I\big\{\phi_0({\bf Z})>0\big\},
\end{equation*}
where ${\bf X}^{(\delta)} = I\big\{\delta=1\big\}{\bf X}+I\big\{\delta=2\big\}{\bf Z}$ represents the observable covariate, then it follows from our Theorem \ref{B1} below that the classifier $\Gamma^{\mbox{\tiny B}}$ has the lowest misclassification error.
In the more general setting with M missing patterns, if we let $X_j^{(k)}= \langle {\boldsymbol\chi},\psi_j\rangle_{s_k}$ then, with $s_1 :=\mathcal{I}$, we have the vectors of scores 
\begin{equation*}
{\bf X}^{(k)}=(X_1^{(k)}, X_2^{(k)}, \dots) = \big(\langle {\boldsymbol\chi}, \psi_1\rangle_{s_k}, \langle {\boldsymbol\chi}, \psi_2\rangle_{s_k}, \dots\dots\big),~k=1,\dots,M.
\end{equation*}
Clearly, when $\delta=k$, we only observe ${\bf X}^{(k)}$ in which case a classifier is any function of the form $g_k: \ell_2\to \{0,1\}.$ Hence,  any classifier can be written in the general form 
\begin{eqnarray}\label{GEN}
\Gamma({\bf X}^{(\delta)})= \sum_{k=1}^M I\{\delta=k\}\cdot g_k({\bf X}^{(k)}), ~~\mbox{where}~~ {\bf X}^{(\delta)} := \sum_{k=1}^M {\bf X}^{(k)} I\{\delta=k\}. 
\end{eqnarray}
Now, let 
\begin{equation}\label{E8}
\phi_k({\bf X}^{(k)}) = E\left[(2Y-1) I\{\delta=k\}\,\big|\, {\bf X}^{(k)}\right],~k=1,\dots,M,
\end{equation}
and define the following classifier (which can be viewed as  the counterpart of (\ref{E4}) on $\ell_2$)
\begin{equation}\label{E9}
\Gamma^{\mbox{\tiny B}}
({\bf X}^{(\delta)})= \sum_{k=1}^M I\{\delta=k\}\cdot I\{\phi_k({\bf X}^k)>0\}.
\end{equation}
Then part (i) of the following result shows that the classifier in (\ref{E9}) is optimal.
\begin{thm}\label{B1} 
Let $\Gamma^{\mbox{\tiny B}}$ be the classifier given by (\ref{E9}). Then
	
\vspace{2mm}
{\bf (i)} The classifier $\Gamma^{\mbox{\tiny B}}$ has the lowest misclassification error, i.e., for any other classifier $\Gamma$, one has 
	$\mathbb{P}\{\Gamma^{\mbox{\tiny B}}({\bf X}^{(\delta)})\neq Y\} \leq \mathbb{P}\{\Gamma({\bf X}^{(\delta)})\neq Y\}$.
	
\vspace{2mm}
{\bf (ii)}	The misclassification error of the optimal classifier based on the whole curve is the same as that of the optimal classifier based on the filtered curve, i.e., 	$\mathbb{P}\{\Gamma_0^{\mbox{\tiny B}}(\boldsymbol{\chi}^{(\delta)})\neq Y\}=\mathbb{P}\{\Gamma^{\mbox{\tiny B}}({\bf X}^{(\delta)}) \neq Y\}$, where $\Gamma^{\mbox{\tiny B}}({\bf X}^{(\delta)})$ and $\Gamma_0^{\mbox{\tiny B}}(\boldsymbol{\chi}^{(\delta)})$ are as in (\ref{E9}) and (\ref{E4}), respectively.

\vspace{2mm}
{\bf (iii)}	Let $\Gamma$ be any classifier of the form $\Gamma({\bf X}^{(\delta)})= \sum_{k=1}^M I\{\delta=k\}\cdot I\{\varphi_k({\bf X}^{(k)})>0\}$ for some functions $\varphi_k: \ell_2 \to [-1, 1],~k=1,\dots,M.$ Then $
\mathbb{P}\big\{\Gamma({\bf X}^{(\delta)})\neq Y\big\}-\mathbb{P}\big\{\Gamma^{\mbox{\tiny B}}({\bf X}^{(\delta)})\neq Y\big\}  \leq \sum_{k=1}^M \mathbb{E}\big|\phi_k({\bf X}^{(k)})-\varphi_k({\bf X}^{(k)})\big|$, where $\phi_k({\bf X}^{(k)})$ is as in (\ref{E8}).
\end{thm}
%\underline{The Proof of} Part (i) is the same as that of Theorem \ref{A} whereas the proof of part (ii) is similar to that of Lemma 1 of Mojirsheibani and Shaw (2018).
\begin{rem}\label{REM-1}
{\rm Part (iii) of Theorem \ref{B1} provides a useful tool to bound the difference between the two misclassification errors in terms of the difference between $\phi_k({\bf X}^{(k)})$ that appears in (\ref{E8})  and the function $\varphi_k({\bf X}^{(k)})$. Here, one can think of $\varphi_k({\bf X}^{(k)})$ as an approximation to the unknown function $\phi_k({\bf X}^{(k)})= \mathbb{E}\left[(2Y-1) I\{\delta=k\}\,\big|\, {\bf X}^{(k)}\right]$. Part (ii) of the theorem, which states that the error of the optimal classifier on $L^2$ is the same as that of the optimal classifier in $\ell_2$, is rather intuitive.}
\end{rem}
\subsection{Reduction to finite dimensions and the proposed classifier}\label{MAIN}
Since working in $\ell_2$ is not convenient from a practical point of view, in what follows we consider finite-dimensional versions of the classifier $\Gamma^{\mbox{\tiny B}}$ defined in (\ref{E9}) where ${\bf X}^{(k)}$ will be replaced by the $d$-dimensional vector ${\bf X}^{(d,k)}=(X_1^{(k)},\dots, X_d^{(k)}) = (\langle\boldsymbol{\chi} , \psi_1\rangle_{s_k}, \dots, \langle\boldsymbol{\chi} , \psi_d\rangle_{s_k})$, $k=1,\dots, M$,  (a data-driven choice of the parameter $d$ is discussed later in this section). More specifically, define the function $\phi_{d,k}: \mathbb{R}^d \to [-1, 1]$ by 
\begin{equation}\label{E8B}
\phi_{d,k}({\bf X}^{(d,k)})= \mathbb{E}\left[(2Y-1) I\{\delta=k\}\,\big|\, {\bf X}^{(d,k)}\right]=E\left[(2Y-1) I\{\delta=k\}\,\big|\, X_1^{(k)},\dots, X_d^{(k)}\right],
\end{equation}
$k=1,\dots,M,$ and consider the following version of the classifier in (\ref{E9})
\begin{equation}\label{E9B}
\Gamma^{\mbox{\tiny B},d}({\bf X}^{(d,\delta)}) = \sum_{k=1}^M I\{\delta=k\}\cdot I\left\{\phi_{d,k}({\bf X}^{(d,k)})>0\right\}.
\end{equation}
Here, ${\bf X}^{(d,\delta)}=\sum_{k=1}^M{\bf X}^{(d,k)}\cdot I\{\delta=k\}$. The following result  shows that the classifier $\Gamma^{\mbox{\tiny B},d}$ is optimal:
\begin{thm}\label{B2}
Let $\Gamma^{\mbox{\tiny B},d}$ be the classifier in (\ref{E9B}). Then for any other classifier $\Gamma$ we have 
$\mathbb{P}\{\Gamma^{\mbox{\tiny B},d}({\bf X}^{(d,\delta)})$ $\neq Y\} \leq \mathbb{P}\{\Gamma({\bf X}^{(d,\delta)}) \neq Y\}.$
\end{thm}
The fact that all distributions are unknown implies that the classifier in (\ref{E9B}) is not available in practice and has to be constructed based on the available data. Here we propose a kernel-type methodology. To construct our kernel classifier, we also employ the following data-splitting approach which is in the spirit of the method proposed by Biau et al (2005) in the case of functional nearest neighbor classification (without any missing data).  Let ${\bf X}^{(\delta)}$ be as in (\ref{GEN}) and start by randomly splitting the data $\mathbb{D}_n=\{({\bf X}_1^{(\delta_1)}, Y_1, \delta_1),\dots, ({\bf X}_n^{(\delta_n)},$  $Y_n, \delta_n)\}$ into a {\it training} sample $\mathbb{D}_m$ of size $m$ and a {\it testing sequence} $\mathbb{D}_{\ell}$ of size $\ell=n-m$. Here, $m$ and $\ell$ typically depend on $n$ (they grow with $n$). Next, put
\begin{equation}
\widehat{\phi}_{m,d,\,h_k}({\bf X}^{(d,k)}) = \sum_{i:~ ({\bf X}^{(\delta_i)},Y_i,\delta_i) \in \mathbb{D}_m}(2Y_i-1) I\{\delta_i=k\}\mathcal{K}_k\left(\frac{{\bf X}^{(d,k)}-{\bf X}_i^{(d,k)}}{h_k}\right)\,, ~~~~~\label{E10}
\end{equation}
where ${\bf X}^{(d,k)}$ and ${\bf X}_i^{(d,k)}$ represent the first $d$ components of ${\bf X}^{(k)}$ and ${\bf X}_i^{(k)}$, respectively, and where $\mathcal{K}_k: \mathbb{R}^d\to \mathbb{R}_+$ is the kernel used with the smoothing parameter $h_k$, and define the kernel-type classifier 
\begin{equation}
\Gamma_m^{d}({\bf X}^{(d,\delta)})=\sum_{k=1}^M I\{\delta=k\}I\left\{\widehat{\phi}_{m,d,h_k}({\bf X}^{(d,k)})>0\right\}.   \label{E11}
\end{equation}
Let, $\mathcal{H}\equiv \mathcal{H}_n$ be a grid of positive values from which $h_1,\dots h_M$ are to be selected, and define $\widehat{d}$ and $\widehat{h}_k$ to be the empirically chosen values of $d$ and $h_k$, $k=1,\dots, M$, based on the testing sequence $\mathbb{D}_{\ell}$, i.e., 
\begin{equation}
(\widehat{d},\widehat{h}_1,\dots, \widehat{h}_M) = \argmin_{1\leq d \leq d_n,\, h_k\in \mathcal{H}_n,\, k=1,\dots, M~}\ell^{-1} \sum_{i:~ {(\bf X}^{(\delta_i)}, Y_i,\delta_i) \in \mathbb{D}_{\ell}} I\big\{\Omega_i(m,d,h_1,\dots, h_M)\big\},        \label{E12}
\end{equation}
where the set $\Omega_i$ is given by
\begin{equation} \label{E13}
\Omega_i(m,d,h_1,\dots, h_M) = \left\{\sum_{k=1}^M I\{\delta_i = k\} \cdot I\left\{\widehat{\phi}_{m,d,h_k}({\bf X}_i^{(d,k)})>0\right\}\neq Y_i\right\},
\end{equation} 
and  where $d_n$ diverges with $n$ but not too rapidly (see Remark \ref{REM-2}). The final classifier is then the plug-in version of (\ref{E11}) given by 
\begin{equation}\label{E14}
\widehat{\Gamma}_n({\bf X}^{(\widehat{d},\delta)})=\sum_{k=1}^M I\{\delta=k\}I\left\{\widehat{\phi}_{n,\widehat{d},\widehat{h}_k}({\bf X}^{(\widehat{d},k)})>0\right\}, 
\end{equation}
where the subscript $n$ used in (\ref{E14}) indicates that it is constructed based on the entire data of size $n$. How good is the classifier $\widehat{\Gamma}_n$ in (\ref{E14})? The next theorem shows that under rather minimal assumptions $\widehat{\Gamma}_n$ is strongly optimal, i.e., 
$\mathbb{P}\{\widehat{\Gamma}_n({\bf X}^{(\widehat{d},\delta)})\neq Y | \mathbb{D}_n\} \to^{\mbox{\small a.s.}}~ \mathbb{P}\{\Gamma^{\mbox{\tiny B}}({\bf X}^{(\delta)})\neq Y\},$ as $n\to\infty$.
To present our main results, we first state the following assumption on the kernels used in (\ref{E10}).

\vspace{4mm}
\noindent
{\it Assumption (K)}. \\
The kernel $\mathcal{K}_k$ used in (\ref{E10}) is {\it regular:} A nonnegative kernel $\mathcal{K}$ is said to be regular if there are positive constants $b>0$ and $r>0$ for which $\mathcal{K}(\textbf{x})\geq bI\{ \textbf{x}\in S_{0,r}\}$ and $\int \sup_{\textbf{y}\in \textbf{x}+ S_{0,r}}\mathcal{K}(\textbf{y})d\textbf{x}<\infty,$ where $S_{0,r}$ is the ball of radius $r$ centered at the origin. (For more on regular kernels see, for example, Gy\"{o}rfi et al (2002).)

\vspace{4mm}
\begin{thm}\label{C}
	Suppose that Assumption (K) holds.
	Also assume that, as $n\to\infty$, we have  $\ell\equiv \ell(n)\to \infty$, $m\equiv m(n)\to \infty$, $\ell^{-1} \log |\mathcal{H}_n| \to 0$, and $\ell^{-1} \log d_n \to 0$, where $|\mathcal{H}_n|$ is the cardinality of the set $\mathcal{H}_n$. Suppose that for each $k=1,\dots,M$, there is an $h_k\equiv h_k(n) \in \mathcal{H}_n$ such that $\max_{1\leq k \leq M} h_k\to 0$ and $\min_{1\leq k\leq M} m h_k^{d_n}\to \infty$, as $n \to\infty$. 	Then the classifier $\widehat{\Gamma}_n$ is asymptotically strongly optimal, i.e., 
	$$
	\mathbb{P}\left\{ \widehat{\Gamma}_n({\bf X}^{(\widehat{d},\delta)})\neq Y\,\big|\,\mathbb{D}_n\right\} \longrightarrow^{\mbox{\rm \small a.s.}}~
	\mathbb{P}\left\{\Gamma^{\mbox{\tiny B}}({\bf X}^{(\delta)})\neq Y\right\},
	$$
	as $n\to\infty$, where $\Gamma^{\mbox{\tiny B}}$ is the theoretically optimal classifier appearing in Theorem \ref{B1}.
\end{thm}
\begin{rem}\label{REM-2}{\rm 
	The conditions imposed on $h_k\equiv h_k(n)$ and $d_n$ in the statement of Theorem \ref{C} are satisfied if $d_n$ does not grow too rapidly and $h_k$ converges to zero slowly, as $n\to \infty$. In fact, if we take $d_n=(\log n^{c_0})^{1-\gamma}$ for any $c_0>0$ and any $0<\gamma< 1$, and if, for example, $h_k=(\log n^{c_k})^{-1}$ for any $c_k>0$, then it is straightforward to see that $m h_k^{d_n}\to \infty,$ as $n\to \infty$. Intuitively, the slow rate of convergence (logarithmic) of $h_k$ to zero is not necessarily unrealistic here and, in a sense, can be tied to the increasing dimension $d_n$. In fact, in what Ferraty and Vieu (2006; page 211) refer to as the {\it curse of infinite dimensionality}, the authors argue that in the problem of kernel regression estimation for the general regression function $\mathbb{E}[Y|{\bf X}={\bf x}]$ with a functional covariate ${\bf X}$, the smoothing parameter $h\equiv h(n)$ can be of order $(\log n)^u$ for some $u<0$. } 
\end{rem} 

%%%%%%%%%%%%%%%%%%%%%%%%%%%%%%%%%%%%%%%%%%%%%%%%%%%%%%%%%%%%%%%%%%%%%%%%%%%%%%%%
%%%%%%%%%%%%%%%%%%%%%%%%%%%%%%%%%%%%%%%%%%%%%%%%%%%%%%%%%%%%%%%%%%%%%%%%%%%%%%%%
%%%%%%%%%%%%%%%%%   Numerical Examples Begin:  %%%%%%%%%%%%%%%%%%%%%%%%%%%%%%%%%%
%%%%%%%%%%%%%%%%%%%%%%%%%%%%%%%%%%%%%%%%%%%%%%%%%%%%%%%%%%%%%%%%%%%%%%%%%%%%%%%%
%%%%%%%%%%%%%%%%%%%%%%%%%%%%%%%%%%%%%%%%%%%%%%%%%%%%%%%%%%%%%%%%%%%%%%%%%%%%%%%%

\subsection{Numerical examples}\label{Example}
\subsubsection{Simulated Data}
\begin{figure}[h!]
	\centering
	\includegraphics[scale = 0.2]{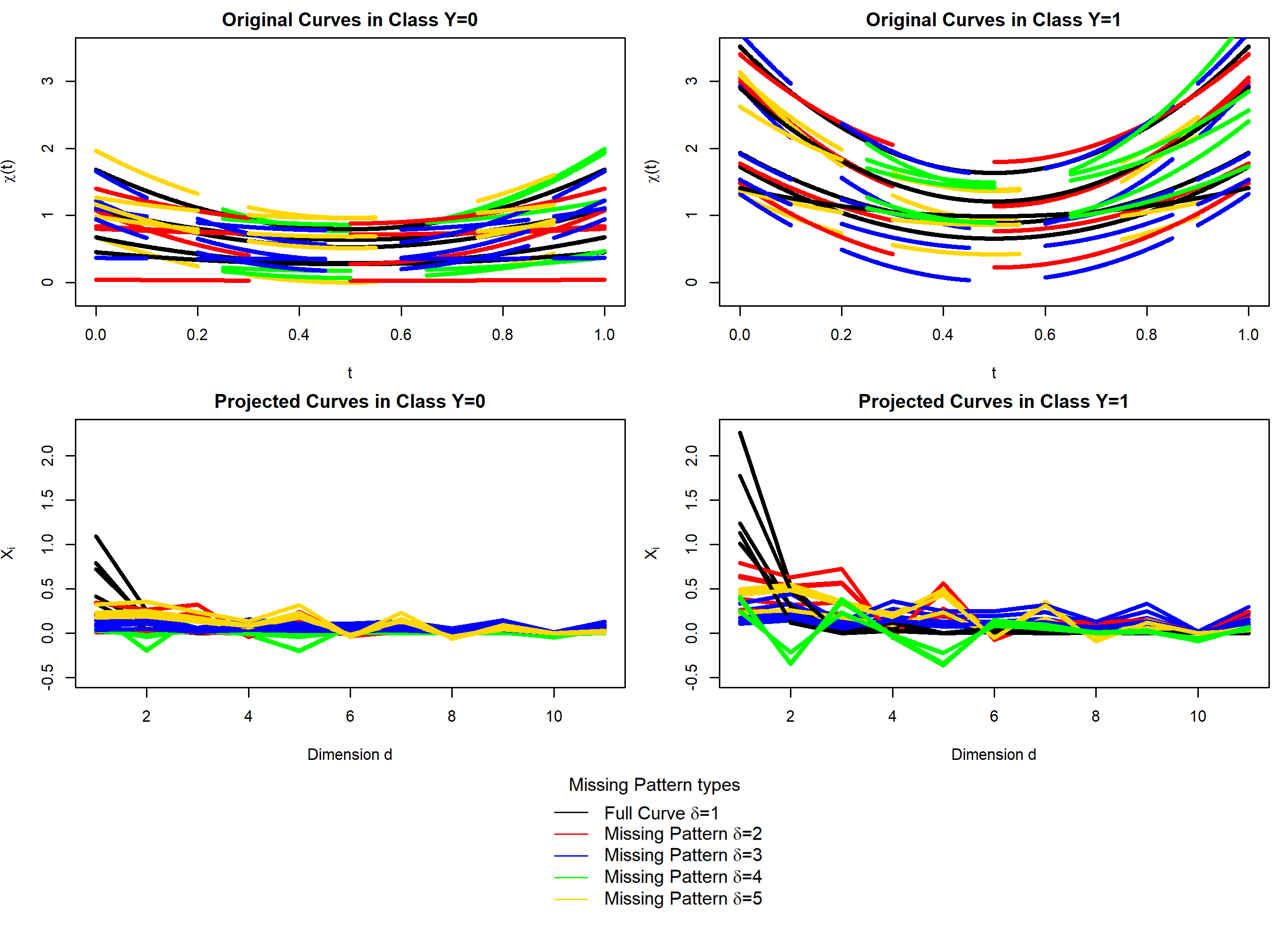}
	\captionsetup{singlelinecheck = false, justification=justified, font = footnotesize}
	\caption{\fontsize{10}{4}\selectfont A sample of simulated curves  with their projections. Here, 30\% of the data contain some unobserved fragments}
	\label{fig:samplecurves}
\end{figure}

Here, we provide some numerical examples to assess the performance of the methods proposed in the previous section. In this analysis, we develop classifiers to predict the unknown class $Y=0$ or $Y=1$ of a functional covariate $\boldsymbol{\chi}(t)$, defined in $L^2([0,1])$, that may have missing fragments. Adopting the missing pattern setup of Section \ref{back}, without loss of generality let $s_1 := \mathcal{I}= [0,1]$. Also, let  $s_2 = [0,0.3]\cup[0.5,1] \subset\mathcal{I}$, 
$s_3 = [0,0.1]\cup[0.2,0.45]\cup[0.6,0.85]\cup[0.9,1] \subset\mathcal{I}$, $s_4 = [0.25,0.5]\cup[0.65,1] \subset\mathcal{I}$, and $s_5 = [0,0.2]\cup[0.3,0.55]\cup[0.75,0.9] \subset\mathcal{I}$. We consider two cases of  missing patterns: $M = 3$ and $M = 5$. In the case of $M=3$, the patterns used are $s_1$, $s_2$, and $s_3$. Next, samples of functional observations $\big(\boldsymbol{\chi}^{(\delta_i)}_i, Y_i, \delta_i\big),~i=1,\dots,n$, are generated based on rules which are similar to the approach of Rachdi and View (2007) and  Mojirsheibani and Shaw (2018) as follows:
\[
\boldsymbol{\chi}_i(t) = A_i(t-0.5)^2 + B_i, ~~~\, i=1,2,\dots,n
\]
where $t \in s =s_1,s_2, s_3, s_4$, or $s_5$ depending on whether $\delta_i = 1,2,3,4$ or $5$.  Regarding the values of $A_i$ and $B_i$, if $Y_i = 1$ then $A_i\overset{\mbox{\tiny iid}}{\sim} N(5,2^2)$ and $B_i \overset{\mbox{\tiny iid}}{\sim}N(1,0.5^2)$, otherwise if $Y_i = 0$ then $A_i\overset{\mbox{\tiny iid}}{\sim} \mbox{Unif}\,(0,5)$ and $B_i \overset{\mbox{\tiny iid}}{\sim} \mbox{Unif}\,(0,1)$. The class probabilities are taken to be $P(Y=1)=0.5 =P(Y=0)$. With respect to
the missing probability mechanism, we consider a logistic-type model
\begin{equation} \label{sim:eq1}
P\big\{\delta = 1 \,\big|Y = y, \mbox{$\boldsymbol{\chi}$}(t) = \chi(t), t \in \mathcal{I}\big\} \\
= 
\frac{\exp\Big\{a(1-y) + b\int_{s}\chi(t)\,dt + c\int_{[0,1]\setminus s}t\cdot
	\chi(t)\,dt \Big\}}{1+\exp\Big\{a(1-y) + b\int_{s}\chi(t)\,dt + c\int_{[0,1]\setminus s}t\cdot\chi(t)\,dt \Big\}},
\end{equation}
where the set $s$ can be selected to be any one of the missing patterns $s_k$, $k=2,\dots, M$, with probability $1/(M-1)$. The coefficients $a ,b,$ and $c$ in (\ref{sim:eq1}) can be adjusted to control the missing data rate. They can also be adjusted to control the level of dependency of the missing probability in (\ref{sim:eq1}) on $Y$ and on the observed and unobserved segments of the curve.
As for the choice of the basis functions, we used the  Fourier basis $\big\{\psi_1(t)= 1,\, \psi_{2k}(t)= \sqrt{2}\cos(\pi kt),$ $\psi_{2k+1}(t)= \sqrt{2}\sin(2\pi kt),\, k\geq 1\big\}$ which forms a complete orthonormal basis for $L^2([0,1])$; see, for example,  Zygmund (1959) and Sansone (1969). 
%There are also other choices such as $\big\{\psi_{k}(t)= \sqrt{2}\sin(\pi kt),\, k\geq 1\big\}$ as well as $\big\{\psi_0(t)= 1,\, \psi_{k}(t)= \sqrt{2}\cos(\pi kt),\, k\geq 1\big\}$
%
%
%
%
%
%
%The functional covariates $\big\{\boldsymbol{\chi}^{(\delta_i)}_i(t)\big\}_{i=1}^n$ are then replaced by $d_n$-dim vectors $\big\{\boldsymbol{X}^{(d_n,\delta_i)}_i\big\}_{i=1}^n$ as discussed in the notation of Section \ref{MAIN}, where we have taken  $d_n \approx 2.5\ln (n) $ (for details and justification, see Remark \ref{REM-2} Section \ref{MAIN}). 
Figure \ref{fig:samplecurves} shows a few realizations of the simulated curves $\boldsymbol{\chi}|_{s_k}$  as well as their corresponding $d$-dim projections, $\boldsymbol{X}^{(d,k)}$, for $d=1,\dots, 10$
and $k=1,\dots,5$. \\

\begin{figure}[h!]
		\centering 
		\includegraphics[width=\textwidth,height = 12cm]{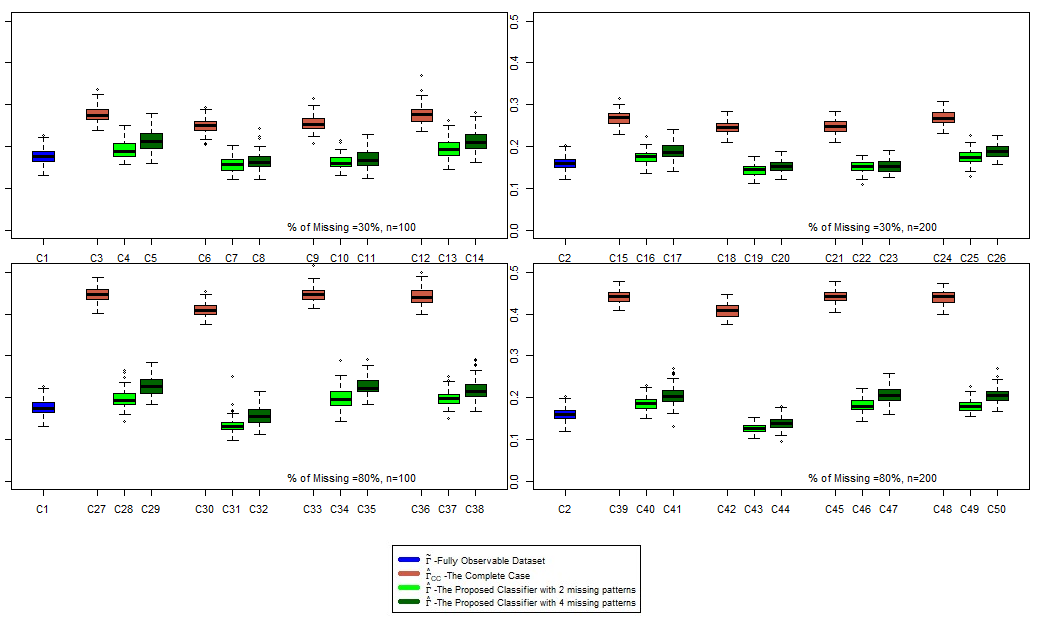}
		\captionsetup{singlelinecheck = false, justification=justified, font = footnotesize}
		\caption{\fontsize{10}{4}\selectfont Boxplots of the error rates of all classifiers (C1, C2, $\dots,$ C50) that appear in Table \ref{tab:result} }
		\label{fig:boxplot}
	\end{figure}

\noindent
Next, we constructed our proposed classifier $\widehat{\Gamma}_n$, given by  (\ref{E14}), based on two different sample sizes, $n = 100$ and $n = 200$, as well as several choices for the constants $a,b,$ and $c$ in (\ref{sim:eq1}) for each of the missing patterns. The parameters $h_{k}$ and $d$ were selected using a data splitting approach (with a splitting ratio of 65:35 for the training set to the testing sequence)  from a grid of equally spaced values $h$ in $[0,1]$ and $1\leq d \leq d_n$, based on the procedure in (\ref{E11}) and  (\ref{E12}) with Gaussian kernels. Here, we took $d_n \approx 2.5 \ln(n)$; see Remark \ref{REM-2} for details and the justification for the choice of $d_n$. This process was repeated for 20 such random sample splits and  the values of $h_k$ and $d$ that minimized the average error were selected; these are denoted by $\widehat{h}_k$ and $\widehat{d}$ which appear in (\ref{E12}). 
In addition to the proposed classifier $\widehat{\Gamma}_n$, we also constructed the classifier based on the complete case analysis, which will be denoted by $\widehat{\Gamma}_{CC}$, (it uses complete cases only) as well as the classifier corresponding to the case with no missing data (i.e., when all covariates are fully observable), to be denoted by $\widetilde{\Gamma}_n$, which was proposed by Biau  et al. (2005). Furthermore, our analysis here includes different missingness mechanisms such as the ``Not Missing At Random'' (NMAR), the ``Missing At Random'' (MAR), and the ``Missing Completely At Random'' (MCAR) scenarios.
These classifiers are then used to classifying 1000 additional observations from the same underlying distribution
of the data. The entire above process was repeated a total of 100 times and the average misclassification errors (over 100 Monte Carlo runs) were computed.  Our findings are summarized in Table \ref{tab:result}. The constants $a, b, c$ (of equation (\ref{sim:eq1})) corresponding to pattern $s_2$ are reported in columns $a_2,b_2,c_2$ of Table \ref{tab:result}, those corresponding to $s_3$ are reported in columns $a_3,b_3,c_3$, and so on. The numbers appearing in parentheses are the standard errors of the reported misclassification errors. 
Figure \ref{fig:boxplot} provides boxplots of the error rates of various classifiers. 
As shown in Table \ref{tab:result} and Figure \ref{fig:boxplot}, for both sample sizes, the error rate of the classifier $\widehat{\Gamma}_n$ is lower than that of $\widehat{\Gamma}_{CC}$ regardless of the missingness mechanism  or the number of missing patterns involved. This is particularly true  when the percentage of missing data is at 80\%.  In passing, we note that the proposed classifier $\widehat{\Gamma}_n$ can also perform better than $\widetilde{\Gamma}_n$ whenever the dependence of the missing probability mechanism on class $Y$ (as defined via (\ref{sim:eq1})) dominates its dependence on the observed and/or unobserved segments of the curves (i.e., the constant $a$ is orders of magnitude larger than $b$ and $c$ in (\ref{sim:eq1})); see, for example, the cases C7, C8, C22, C23, C31, C32, C43, C44 in Table \ref{tab:result}. This shows that in such cases the variable $\delta$ can sometimes be a much  better predictor of the class variable $Y$ than the missing fragments of the covariate curves.

% Table generated by Excel2LaTeX from sheet 'Sheet4'
\begin{table}[htbp]
	\centering
	\captionsetup{singlelinecheck = false, justification=justified, font = footnotesize}
	\caption{\fontsize{10}{4}\selectfont Misclassification errors of 
		 $\widetilde{\Gamma}_n$ (fully observable data), $\widehat{\Gamma}_{CC}$ (complete case analysis), $\widehat{\Gamma}_n$ (the proposed classifier with $M=3$ and $5$ missing patterns). The numbers in parentheses are the standard errors over 100 Monte Carlo runs.}
	\resizebox{\textwidth}{!}{ \begin{tabular}{|l|r|p{5.285em}|rp{2.855em}r|rp{2.855em}r|rp{2.855em}r|rp{2.855em}r|l|p{4.215em}|p{4.215em}|p{4.215em}|}
			\toprule
			\multicolumn{1}{|p{4em}|}{\textbf{\% of Missing }} & \multicolumn{1}{p{2.855em}|}{\textbf{n}} & \textbf{Missingness Mechanism} & \multicolumn{1}{p{2.855em}}{\textbf{a2}} & \textbf{b2} & \multicolumn{1}{p{2.855em}|}{\textbf{c2}} & \multicolumn{1}{p{2.855em}}{\textbf{a3}} & \textbf{b3} & \multicolumn{1}{p{2.855em}|}{\textbf{c3}} & \multicolumn{1}{p{2.855em}}{\textbf{a4}} & \textbf{b4} & \multicolumn{1}{p{2.855em}|}{\textbf{c4}} & \multicolumn{1}{p{2.855em}}{\textbf{a5}} & \textbf{b5} & \multicolumn{1}{p{2.855em}|}{\textbf{c5}} & \multicolumn{1}{p{4.215em}|}{\textbf{Error of $\widetilde{\Gamma}_n$}} & \textbf{Error of $\widehat{\Gamma}_{CC}$} & \textbf{Error of $\widehat{\Gamma}_n$ with $M$=3 missing patterns} & \textbf{Error of $\widehat{\Gamma}_n$ with $M$=5 missing patterns} \\
			\midrule
			&       & NMAR  & \multicolumn{1}{l}{0} & \multicolumn{1}{l}{0.95} & \multicolumn{1}{l|}{0.13} & \multicolumn{1}{l}{0} & \multicolumn{1}{l}{0.9} & \multicolumn{1}{l|}{1} & \multicolumn{1}{l}{0} & \multicolumn{1}{l}{1.05} & \multicolumn{1}{l|}{0.13} & \multicolumn{1}{l}{0} & \multicolumn{1}{l}{1.2} & \multicolumn{1}{l|}{1} &       & C3\newline{}0.2774\newline{}(0.0179) & C4\newline{}0.1920\newline{}(0.0215) & C5\newline{}0.2124\newline{}(0.0241) \\
			\cmidrule{3-15}\cmidrule{17-19}          & \multicolumn{1}{c|}{\multirow{2}[4]{*}{100}} & NMAR  & \multicolumn{1}{l}{2} & \multicolumn{1}{l}{0.01} & \multicolumn{1}{l|}{0.8} & \multicolumn{1}{l}{2} & \multicolumn{1}{l}{0.01} & \multicolumn{1}{l|}{0.4} & \multicolumn{1}{l}{2} & \multicolumn{1}{l}{0.01} & \multicolumn{1}{l|}{0.3} & \multicolumn{1}{l}{2} & \multicolumn{1}{l}{0.01} & \multicolumn{1}{l|}{0.3} & \multicolumn{1}{p{4.215em}|}{C1\newline{}0.1771\newline{}(0.0178)} & C6\newline{}0.2502\newline{}(0.0185) & C7\newline{}0.1581\newline{}(0.0176) & C8\newline{}0.1650\newline{}(0.0199) \\
			\cmidrule{3-15}\cmidrule{17-19}          &       & MAR   & \multicolumn{1}{l}{1.9} & \multicolumn{1}{l}{0.075} & \multicolumn{1}{l|}{0} & \multicolumn{1}{l}{1.9} & \multicolumn{1}{l}{0.075} & \multicolumn{1}{l|}{0} & \multicolumn{1}{l}{2} & \multicolumn{1}{l}{0.085} & \multicolumn{1}{l|}{0} & \multicolumn{1}{l}{2} & \multicolumn{1}{l}{0.085} & \multicolumn{1}{l|}{0} &       & C9\newline{}0.2549\newline{}(0.0187) & C10\newline{}0.1626\newline{}(0.0165) & C11\newline{}0.1700\newline{}(0.0231) \\
			\cmidrule{3-15}\cmidrule{17-19}    \multirow{2}[4]{*}{30\%} &       & MCAR  &       & NA    &       &       & NA    &       &       & NA    &       &       & NA    &       &       & C12\newline{}0.2773\newline{}(0.0227) & C13\newline{}0.1939\newline{}(0.0224) & C14\newline{}0.2121\newline{}(0.0238) \\
			\cmidrule{2-19}          &       & NMAR  & \multicolumn{1}{l}{0} & \multicolumn{1}{l}{0.95} & \multicolumn{1}{l|}{0.13} & \multicolumn{1}{l}{0} & \multicolumn{1}{l}{0.9} & \multicolumn{1}{l|}{1} & \multicolumn{1}{l}{0} & \multicolumn{1}{l}{1.05} & \multicolumn{1}{l|}{0.13} & \multicolumn{1}{l}{0} & \multicolumn{1}{l}{1.2} & \multicolumn{1}{l|}{1} &       & C15\newline{}0.2672\newline{}(0.0159) & C16\newline{}0.1743\newline{}(0.0157) & C17\newline{}0.1883\newline{}(0.0202) \\
			\cmidrule{3-15}\cmidrule{17-19}          & \multicolumn{1}{c|}{\multirow{2}[4]{*}{200}} & NMAR  & \multicolumn{1}{l}{2} & \multicolumn{1}{l}{0.01} & \multicolumn{1}{l|}{0.8} & \multicolumn{1}{l}{2} & \multicolumn{1}{l}{0.01} & \multicolumn{1}{l|}{0.4} & \multicolumn{1}{l}{2} & \multicolumn{1}{l}{0.01} & \multicolumn{1}{l|}{0.3} & \multicolumn{1}{l}{2} & \multicolumn{1}{l}{0.01} & \multicolumn{1}{l|}{0.3} & \multicolumn{1}{p{4.215em}|}{C2\newline{}0.1607\newline{}(0.0154)} & C18\newline{}0.2451\newline{}(0.0137) & C19\newline{}0.1430\newline{}(0.0128) & C20\newline{}0.1517\newline{}(0.0135) \\
			\cmidrule{3-15}\cmidrule{17-19}          &       & MAR   & \multicolumn{1}{l}{1.9} & \multicolumn{1}{l}{0.075} & \multicolumn{1}{l|}{0} & \multicolumn{1}{l}{1.9} & \multicolumn{1}{l}{0.075} & \multicolumn{1}{l|}{0} & \multicolumn{1}{l}{2} & \multicolumn{1}{l}{0.085} & \multicolumn{1}{l|}{0} & \multicolumn{1}{l}{2} & \multicolumn{1}{l}{0.085} & \multicolumn{1}{l|}{0} &       & C21\newline{}0.2470\newline{}(0.0161) & C22\newline{}0.1510\newline{}(0.0147) & C23\newline{}0.1529\newline{}(0.0146) \\
			\cmidrule{3-15}\cmidrule{17-19}          &       & MCAR  &       & NA    &       &       & NA    &       &       & NA    &       &       & NA    &       &       & C24\newline{}0.2686\newline{}(0.0180) & C25\newline{}0.1744\newline{}(0.0171) & C26\newline{}0.1871\newline{}(0.0159) \\
			\midrule
			&       & NMAR  & \multicolumn{1}{l}{0} & \multicolumn{1}{l}{-1.9} & \multicolumn{1}{l|}{1.5} & \multicolumn{1}{l}{0} & \multicolumn{1}{l}{-1.45} & \multicolumn{1}{l|}{-3} & \multicolumn{1}{l}{0} & \multicolumn{1}{l}{-2.1} & \multicolumn{1}{l|}{1.5} & \multicolumn{1}{l}{0} & \multicolumn{1}{l}{-2} & \multicolumn{1}{l|}{-3} &       & C27\newline{}0.4463\newline{}(0.0184) & C28\newline{}0.1970\newline{}(0.0205) & C29\newline{}0.2284\newline{}(0.0240) \\
			\cmidrule{3-15}\cmidrule{17-19}          & \multicolumn{1}{c|}{\multirow{2}[4]{*}{100}} & NMAR  & \multicolumn{1}{l}{-5} & \multicolumn{1}{l}{-0.4} & \multicolumn{1}{l|}{0.25} & \multicolumn{1}{l}{-4} & \multicolumn{1}{l}{-0.5} & \multicolumn{1}{l|}{-0.15} & \multicolumn{1}{l}{-5} & \multicolumn{1}{l}{-0.4} & \multicolumn{1}{l|}{0.25} & \multicolumn{1}{l}{-4} & \multicolumn{1}{l}{-0.5} & \multicolumn{1}{l|}{-0.15} & \multicolumn{1}{p{4.215em}|}{C1\newline{}0.1771\newline{}(0.0178)} & C30\newline{}0.4096\newline{}(0.0158) & C31\newline{}0.1335\newline{}(0.0183) & C32\newline{}0.1554\newline{}(0.0207) \\
			\cmidrule{3-15}\cmidrule{17-19}          &       & MAR   & \multicolumn{1}{l}{1} & \multicolumn{1}{l}{-3} & \multicolumn{1}{l|}{0} & \multicolumn{1}{l}{-1.45} & \multicolumn{1}{l}{-0.95} & \multicolumn{1}{l|}{0} & \multicolumn{1}{l}{0.6} & \multicolumn{1}{l}{-3} & \multicolumn{1}{l|}{0} & \multicolumn{1}{l}{-1.9} & \multicolumn{1}{l}{-0.95} & \multicolumn{1}{l|}{0} &       & C33\newline{}0.4464\newline{}(0.0165) & C34\newline{}0.1991\newline{}(0.0244) & C35\newline{}0.2273\newline{}(0.0202) \\
			\cmidrule{3-15}\cmidrule{17-19}    \multirow{2}[4]{*}{80\%} &       & MCAR  &       & NA    &       &       & NA    &       &       & NA    &       &       & NA    &       &       & C36\newline{}0.4419\newline{}(0.0187) & C37\newline{}0.1978\newline{}(0.0181) & C38\newline{}0.2178\newline{}(0.0256) \\
			\cmidrule{2-19}          &       & NMAR  & \multicolumn{1}{l}{0} & \multicolumn{1}{l}{-1.9} & \multicolumn{1}{l|}{1.5} & \multicolumn{1}{l}{0} & \multicolumn{1}{l}{-1.45} & \multicolumn{1}{l|}{-3} & \multicolumn{1}{l}{0} & \multicolumn{1}{l}{-2.1} & \multicolumn{1}{l|}{1.5} & \multicolumn{1}{l}{0} & \multicolumn{1}{l}{-2} & \multicolumn{1}{l|}{-3} &       & C39\newline{}0.4410\newline{}(0.0157) & C40\newline{}0.1872\newline{}(0.0151) & C41\newline{}0.2048\newline{}(0.0227) \\
			\cmidrule{3-15}\cmidrule{17-19}          & \multicolumn{1}{c|}{\multirow{2}[4]{*}{200}} & NMAR  & \multicolumn{1}{l}{-5} & \multicolumn{1}{l}{-0.4} & \multicolumn{1}{l|}{0.25} & \multicolumn{1}{l}{-4} & \multicolumn{1}{l}{-0.5} & \multicolumn{1}{l|}{-0.15} & \multicolumn{1}{l}{-5} & \multicolumn{1}{l}{-0.4} & \multicolumn{1}{l|}{0.25} & \multicolumn{1}{l}{-4} & \multicolumn{1}{l}{-0.5} & \multicolumn{1}{l|}{-0.15} & \multicolumn{1}{p{4.215em}|}{C2\newline{}0.1607\newline{}(0.0154)} & C42\newline{}0.4087\newline{}(0.0164) & C43\newline{}0.1264\newline{}(0.0118) & C44\newline{}0.1394\newline{}(0.0150) \\
			\cmidrule{3-15}\cmidrule{17-19}          &       & MAR   & \multicolumn{1}{l}{1} & \multicolumn{1}{l}{-3} & \multicolumn{1}{l|}{0} & \multicolumn{1}{l}{-1.45} & \multicolumn{1}{l}{-0.95} & \multicolumn{1}{l|}{0} & \multicolumn{1}{l}{0.6} & \multicolumn{1}{l}{-3} & \multicolumn{1}{l|}{0} & \multicolumn{1}{l}{-1.9} & \multicolumn{1}{l}{-0.95} & \multicolumn{1}{l|}{0} &       & C45\newline{}0.4412\newline{}(0.0153) & C46\newline{}0.1824\newline{}(0.0167) & C47\newline{}0.2061\newline{}(0.0194) \\
			\cmidrule{3-15}\cmidrule{17-19}          &       & MCAR  &       & NA    &       &       & NA    &       &       & NA    &       &       & NA    &       &       & C48\newline{}0.4398\newline{}(0.0179) & C49\newline{}0.1800\newline{}(0.0141) & C50\newline{}0.2058\newline{}(0.0181) \\
			\bottomrule
		\end{tabular}}%
		\label{tab:result}%
	\end{table}%

\vspace{8mm}	
	
%%%%%%%%%%%%%%%%%%%%%%%%%%%%%%%%%%%%%%%%%%%%%%%%%%%%%%%%%%%%%%%%%%%%%%%%%%%%%%%%
%%%%%%%%%%%%%%%%%%%%%%%%%%%%%%%%%%%%%%%%%%%%%%%%%%%%%%%%%%%%%%%%%%%%%%%%%%%%%%%%
%%%%%%%%%%%%%%%%%%%%%%% Three Real Datasets %%%%%%%%%%%%%%%%%%%%%%%%%%%%%%%%%%%%
%%%%%%%%%%%%%%%%%%%%%%%%%%%%%%%%%%%%%%%%%%%%%%%%%%%%%%%%%%%%%%%%%%%%%%%%%%%%%%%%
%%%%%%%%%%%%%%%%%%%%%%%%%%%%%%%%%%%%%%%%%%%%%%%%%%%%%%%%%%%%%%%%%%%%%%%%%%%%%%%%

\newpage

\subsubsection{Three real datasets}	
	In this section, we use three real-world datasets to assess the performance of the proposed classifier. 
	%$\widehat\Gamma_n$ in (12). 
	%against the complete case classifier $\widehat\Gamma_{\scriptscriptstyle CC}$. 
	%and the classifier which utilizes the entire fragmented curve proposed by Mojirsheibani and Shaw (2018) which we will denote as $\widehat\Gamma_{\scriptscriptstyle F}$. 
	In every example that follows, the smoothing parameters $h_{k}$ and $d$ 
	%for $\widehat \Gamma_n$ and $h$ for $\widehat \Gamma_{\scriptscriptstyle CC}$ 
	were selected using the same data splitting approach described in Section 2.3.  
	%The classifier $\widehat\Gamma_{\scriptscriptstyle F}$ requires a similar smoothing parameter $h_{k}$ which was chosen in an analogous fashion.\\  
	Across these examples we see that the proposed classifier consistently performs well 
	%outperforms the complete case classifier 
	regardless of the proportion of fragmented curves.\\

	\begin{figure}[H]
		\centering
		\subfloat[]{{\includegraphics[scale = 0.11]{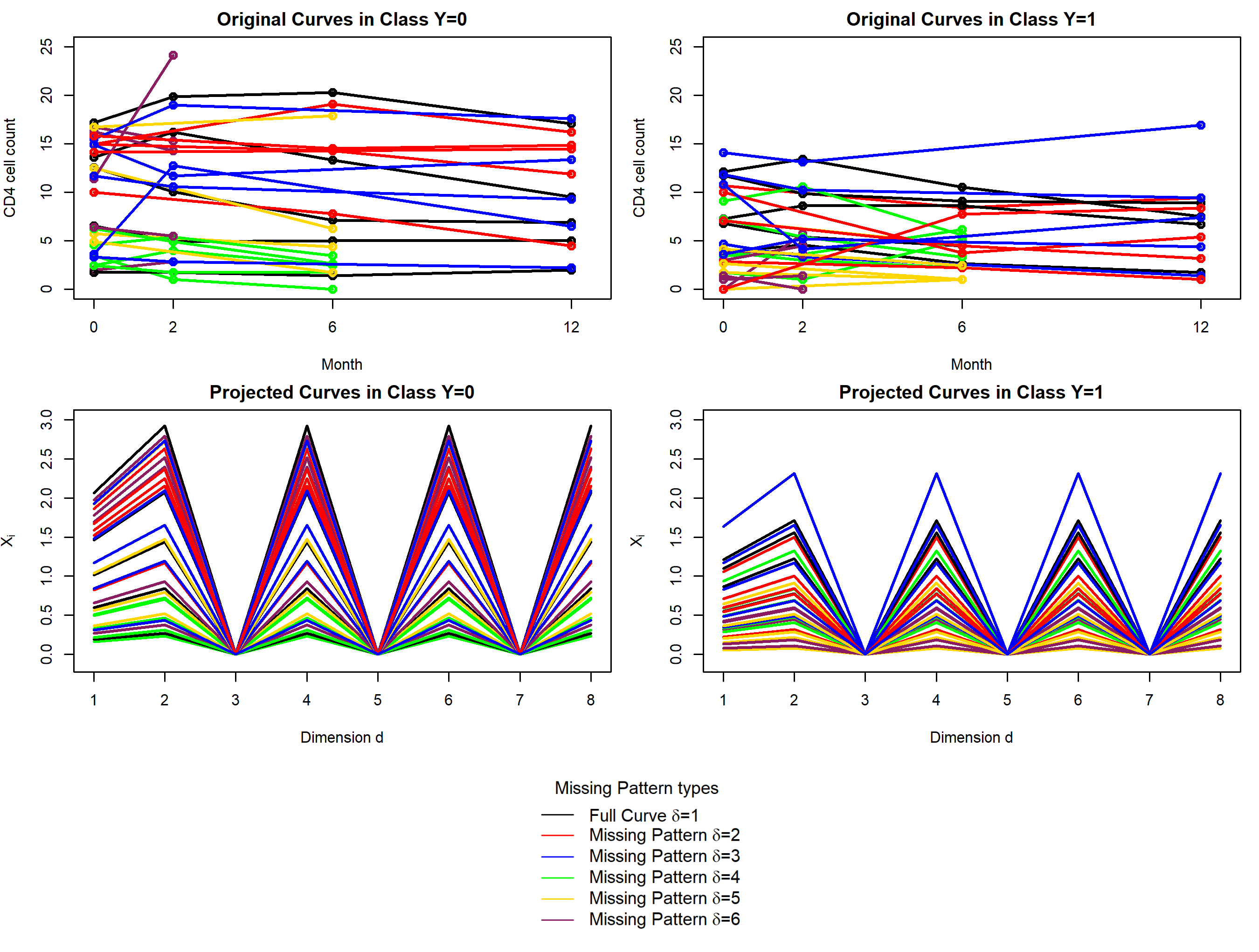}}}%
		\qquad
		\subfloat[]{{\includegraphics[scale = 0.40]{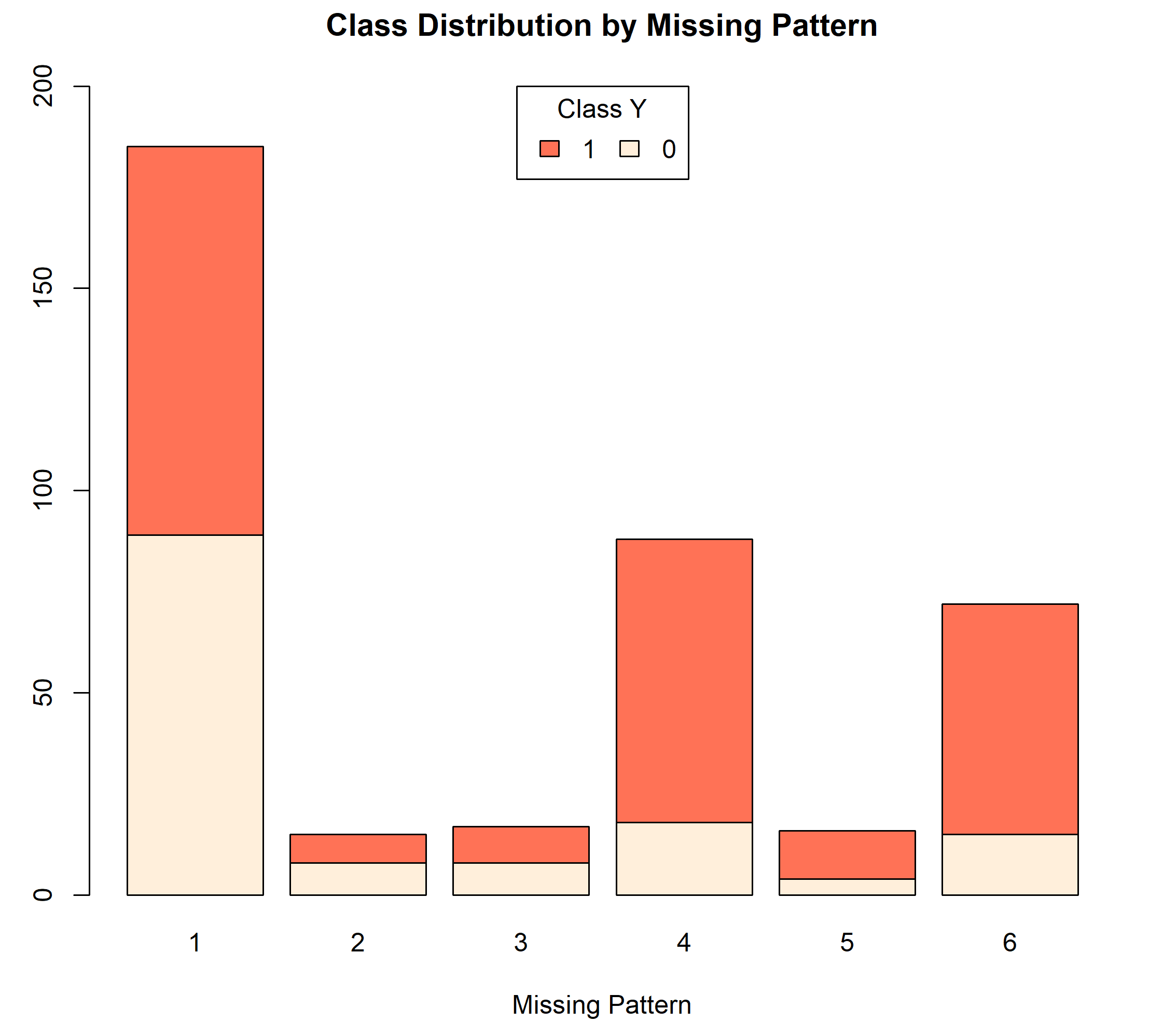}}}%
		\caption{\fontsize{10}{4}\selectfont (a) A sample of curves ${\boldsymbol{\mathcal{X}}}_i$ representing CD4 cell counts measured at each visit for individuals in the HIV treatment study described in Example (A). The curves for those individuals whose data were available for every visit (complete) appear in solid black (in the top two plots) while those whose data were only available for a subset of the visits (missing) are plotted using various colored lines according to their missing pattern.  The corresponding $d$-dimensional vectors of projected curves, $d = 1, \dots, 14$, are displayed below the original curves. (b) Distributions and proportion of class membership for each missing pattern in the utilized subset of data.}%
		\label{aids_curves}%
	\end{figure}
	
	\vspace{4mm}
	{\it Example (A): HIV Treatment, Data from a randomized trial.}
	
	\vspace{1mm}\noindent
	The Community Programs for Clinical Research on AIDS (CPCRA) was a program established by the National Institute of Allergy and Infectious Diseases in 1989.  Seventeen research units were located in 13 cities across the US where the AIDS epidemic was most severe.  Their mission was to expand the clinical research on HIV disease by conducting clinical trials (Cox et al. (1998)).  The aim of one particular trial was to study the efficacy of the drugs Didanosine and Zalcitabine as secondary treatments for patients with HIV who could not tolerate or experienced disease progression despite treatment with Zidovudine, the first developed treatment for HIV. About 400 patients across 133 clinical sites were randomized to receive either Didanosine or Zalcitabine.  The recruitment period lasted one year with follow-up visits at 2, 6, 12 and 18 months after enrollment in the trial (Abrams et al. (1994)).\\ 
	
	AIDS is the most advanced stage of an HIV infection and is diagnosed in the presence of certain opportunistic infections or when an individual's CD4 cell count decreases below 200 cells/$\text{mm}^3$.  An individual with an HIV diagnosis may never develop AIDS, however, expecially if HIV treatments are effective at increasing the individual's CD4 cell count.  In this context, we are interested in whether or not the participant had a previous AIDS diagnosis before entering the trial.  The functional covariate ${\boldsymbol{\mathcal{X}}}_i$ is the repeated measurement of CD4 counts for participants over the duration of the trial. The class variable $y_i$ is coded as 0 = no previous AIDS diagnosis and 1 = previous AIDS diagnosis.\\
	
	The subset of the data used in this example contains 393 observations, 53\% of which are fragmented curves. There are six distinct missing data patterns observed in this dataset including the case where the data are fully observable at all visit times within the first year of the trial.
	%Using the notation of section 2.1, we denote the observed missing patterns as $s_1 := \mathcal{I} = [0, 12], s_2 = \{0\} \cup [6, 12] \subset [0, 12], s_3 = [0, 2] \cup \{12\} \subset [0, 12], s_4 = [0, 2] \cup \{6\} \subset [0, 12], s_5 = [0, 2] \subset [0, 12], s_6 = [0, 6] \subset [0, 12]$. 
	A sample of CD4 cell count curves and corresponding $d$-dimensional vectors of projected curves is displayed in panel (a) of Figure \ref{aids_curves} while the distribution of each missing pattern is displayed in panel (b).\\
	
	We compare the performance of our proposed classifier, $\widehat\Gamma_n$, with that of the classifier based on the complete case analysis, $\widehat \Gamma_{\scriptscriptstyle CC}$. 
	%and the classifier based on the entire fragmented curve, $\widehat \Gamma_{\scriptscriptstyle F}$, 
	To do this, the sample of $n = 393$ individuals was split into a training sequence of size $n' = 196$ and a testing sequence of size $n - n' = 197$. Table \ref{aids_table} provides the average error rates of each classifier committed on the testing sequence over 200 such sample splits with standard errors given in parenthesis as well as a visual display of classifier performance.  Here we see that with over half of the observations being fragemented curves, a complete case classifier eliminates much of the available information and performs poorly compared to the classifier based on the filtered curves.\\ 
	
	\begin{table}[H]
		\captionsetup{singlelinecheck = false, justification=justified,font = footnotesize}
		\centering
		\begin{footnotesize}
			\caption{Error rates for $\widehat \Gamma_n$ (the classifier based on filtered curves) and 
				%$\widehat \Gamma_{\scriptscriptstyle F}$ (the classifier based on the available fragmented curve) and 
				$\widehat \Gamma_{\scriptscriptstyle CC}$ (the complete case analysis).}
			\begin{tabular}{ccccc}
				& & & \hspace{50pt} & \multirow{1}{*}{\includegraphics[scale = 0.35]{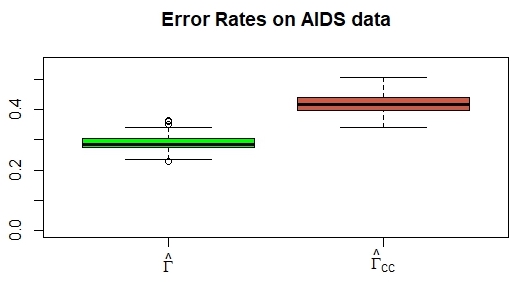}}\\
				\cmidrule[1.25pt]{1-3}
				$\%$ Missing Data & $\widehat \Gamma_n$ & $\widehat \Gamma_{\scriptscriptstyle CC}$ & \hspace{50pt} &\\
				\cmidrule[1.25pt]{1-3}
				$53\%$ & 0.2878 & $0.4187$ & \hspace{50pt} & \\
				& (0.0233) & $(0.0323)$ & \hspace{50pt} & \\
				\cmidrule[1.25pt]{1-3}
				\label{aids_table}
			\end{tabular}
		\end{footnotesize}
	\end{table}
	
	\vspace{100pt}
	
	\begin{figure}[H]%
		\centering
		\subfloat[]{{\includegraphics[scale = 0.12]{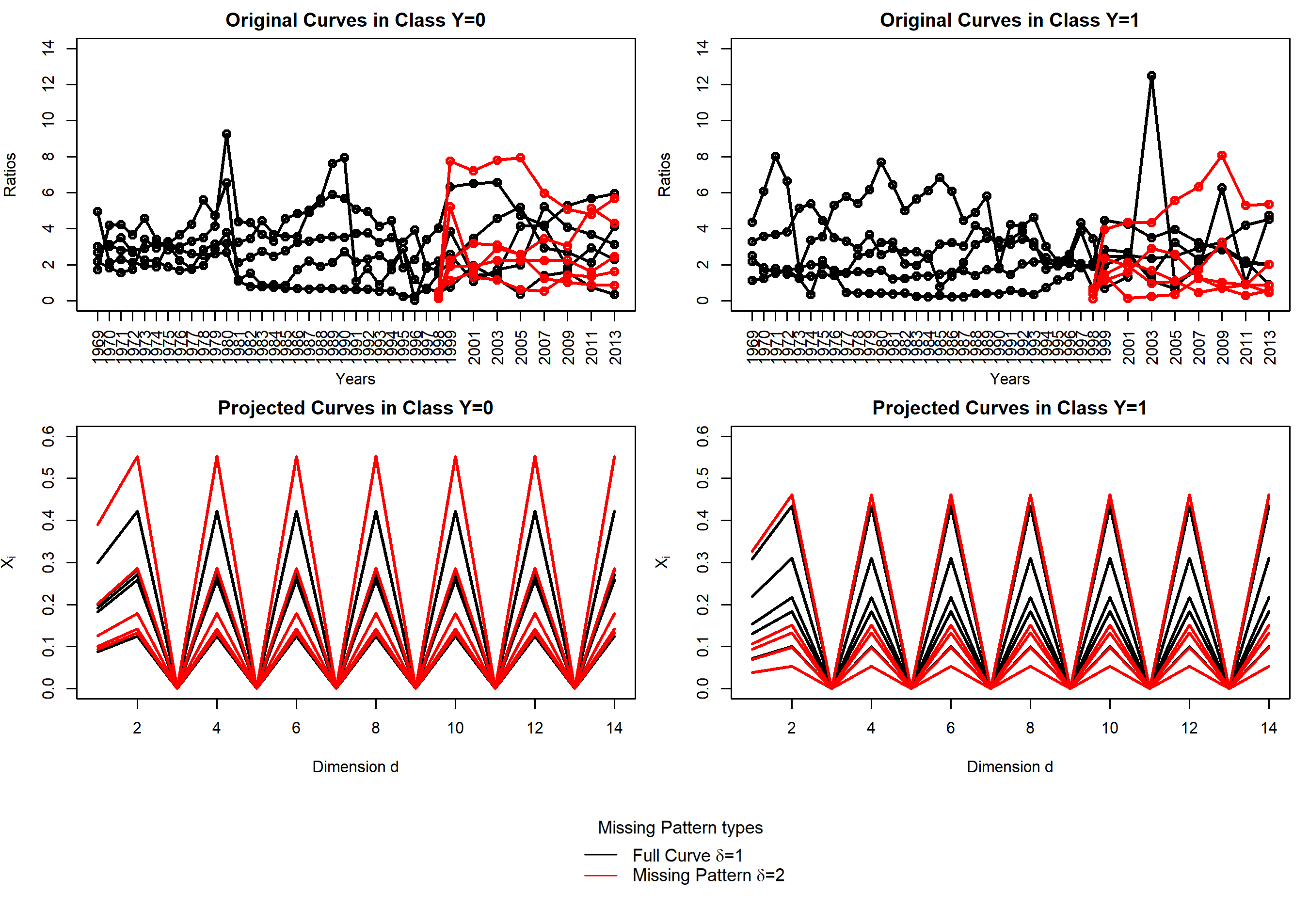}}}%
		\qquad
		\subfloat[]{{\includegraphics[scale = 0.28]{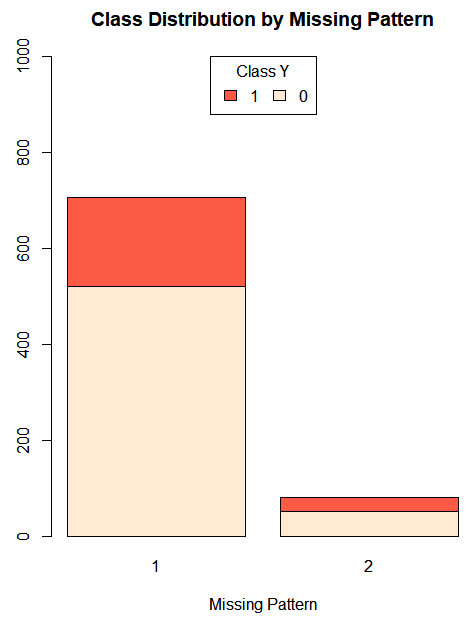}}}%
		\caption{\fontsize{10}{4}\selectfont (a) A sample of curves ${\boldsymbol {\mathcal{X}}}_i$ representing income-to-needs ratios for families participating in the Panel Study of Income Dynamics (PSID), described in Example (B). The curves for those families whose data were available for every survey year (complete) appear in solid black (in the top two plots) while those whose data were available in the subset of years 1998 - 2013 (missing) are plotted in red.  The corresponding $d$-dimensional vectors of projected curves, $d = 1, \dots, 14$, are displayed below the original curves. (b) Distributions and proportion of class membership for each missing pattern in the utilized subset of the data.}%
		\label{PSID_curves}%
	\end{figure}
	
	\vspace{4mm}\noindent
	{\it Example (B):  Panel Study of Income Dynamics.}
	
	\vspace{1mm}\noindent
	The Panel Study of Income Dynamics (PSID)\footnote{The collection of data used in this study was partly supported by the National Institutes of Health under grant number R01 HD069609 and the National Science Foundation under award number 1157698.} is a longitudinal study of socioeconomics and health over lifetimes of participants and across generations of their families. With initial interviews in 1968, follow-up interviews were conducted annualy until 1999 and biennially thereafter. The goal of the PSID is to help researchers understand the complicated dynamics of economic, demographic, health, sociological, and psychological factors. A full description of the dataset is available at {\tt https://psidonline.isr.umich.edu}.  The Child Development Supplement (CDS) was added to the PSID in 1997 and studies a broad array of developmental outcomes in the 0-12 year old children of participants. The dataset used in this example is a subset of the PSID, augmented with the corresponding subset of data from the CDS.\\
	
	It has been shown in the literature that an individual's health status is positively associated with their annual income and that this association originates in childhood (see Brooks-Gunn et al. (1997), Case et al. (2002)).  There is no consensus, however regarding when household income begins affecting a child's health.  This classification problem is based on a study described in Case et al. (2002) which aims to understand the effect that a household's long-run average income has on a child's health.  The functional covariate ${\boldsymbol{ \mathcal{X}}}_i$ is the measure of income-to-needs ratios for participating families over the time period 1969 - 2013. These ratios are calculated by dividing the family's annual household income by the poverty threshold for the corresponding family size (provided by the US Census Bureau).  The class variable $y_i$ is the self-reported health rating of the eldest child in each family, 0 = Excellent - Very Good and 1 = Good - Fair - Poor.\\ 
	
	The subset used in this analysis consisted of $n = 800$ total observations, $87$ ($11\%$) of which were fragmented curves. There are two distinct missing patterns observed in this data, curves that are fully available for the years 1969-2013 and those that are left censored prior to 1998.
	%In the notation of section 2.1, we observed the following patterns: $s_1 = \mathcal I = [1969, 2013]$ and $s_2 = [1998,2013] \subset [1969,2013]$. 
	A small sample of the income to needs ratio curves for families included in the subset of data and their corresponding $d$-dimensional vectors of projected curves is provided in Figure \ref{PSID_curves}.\\ 
	
	We will compare the performance of our proposed classifier, $\widehat\Gamma_n$, with that of the classifier based on the complete case analysis, $\widehat \Gamma_{\scriptscriptstyle CC}$ 
	%and the classifier based on the available fragmented curve, $\widehat \Gamma_{\scriptscriptstyle F}$. 
	To do this, the sample of $n = 800$ families was split into a training sequence of size $n' = 400$ and a testing sequence of size $n - n' = 400$. Table \ref{PSID_table} provides the average error rates of each classifier committed on the testing sequence over 200 such sample splits as well as a visual display of classifier performance. The standard errors are given in parenthesis. In this case we see a marginal gain by classifying using filtered curves (using $\widehat \Gamma_{\scriptscriptstyle n}$) rather than simply performing a complete case analysis.  This is likely due to the small amount of missing data which puts complete case methods on nearly equal footing with other proposed methods.\\ 
	
	\begin{table}[H]
		\captionsetup{singlelinecheck = false, justification=justified,font = footnotesize}
		\centering
		\begin{footnotesize}
			\caption{Error rates for $\widehat \Gamma_n$ (the classifier based on filtered curves) and 
				%$\widehat \Gamma_{\scriptscriptstyle F}$ (the classifier based on the available fragmented curve) and 
				$\widehat \Gamma_{\scriptscriptstyle CC}$ (the complete case analysis).}
			\begin{tabular}{ccccc}
				& & & \hspace{50pt} & \multirow{1}{*}{\includegraphics[scale = 0.375]{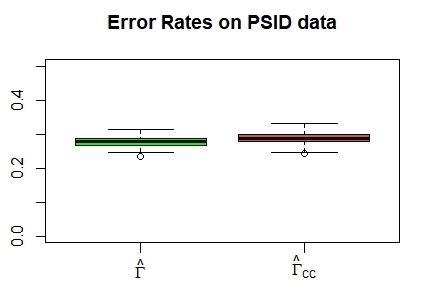}}\\
				\cmidrule[1.25pt]{1-3}
				$\%$ Missing Data & $\widehat \Gamma_n$ & $\widehat \Gamma_{\scriptscriptstyle CC}$ & \hspace{50pt} &\\
				\cmidrule[1.25pt]{1-3}
				$11\%$ & 0.2776 & $0.2894$ & \hspace{50pt} &\\
				& (0.0146) & $(0.0172)$ & \hspace{50pt} &\\
				\cmidrule[1.25pt]{1-3}
				\label{PSID_table}
			\end{tabular}
		\end{footnotesize}
	\end{table}
	
	\vspace{100pt}
	
	\begin{figure}[H]%
		\centering
		\subfloat[]{{\includegraphics[scale = 0.11]{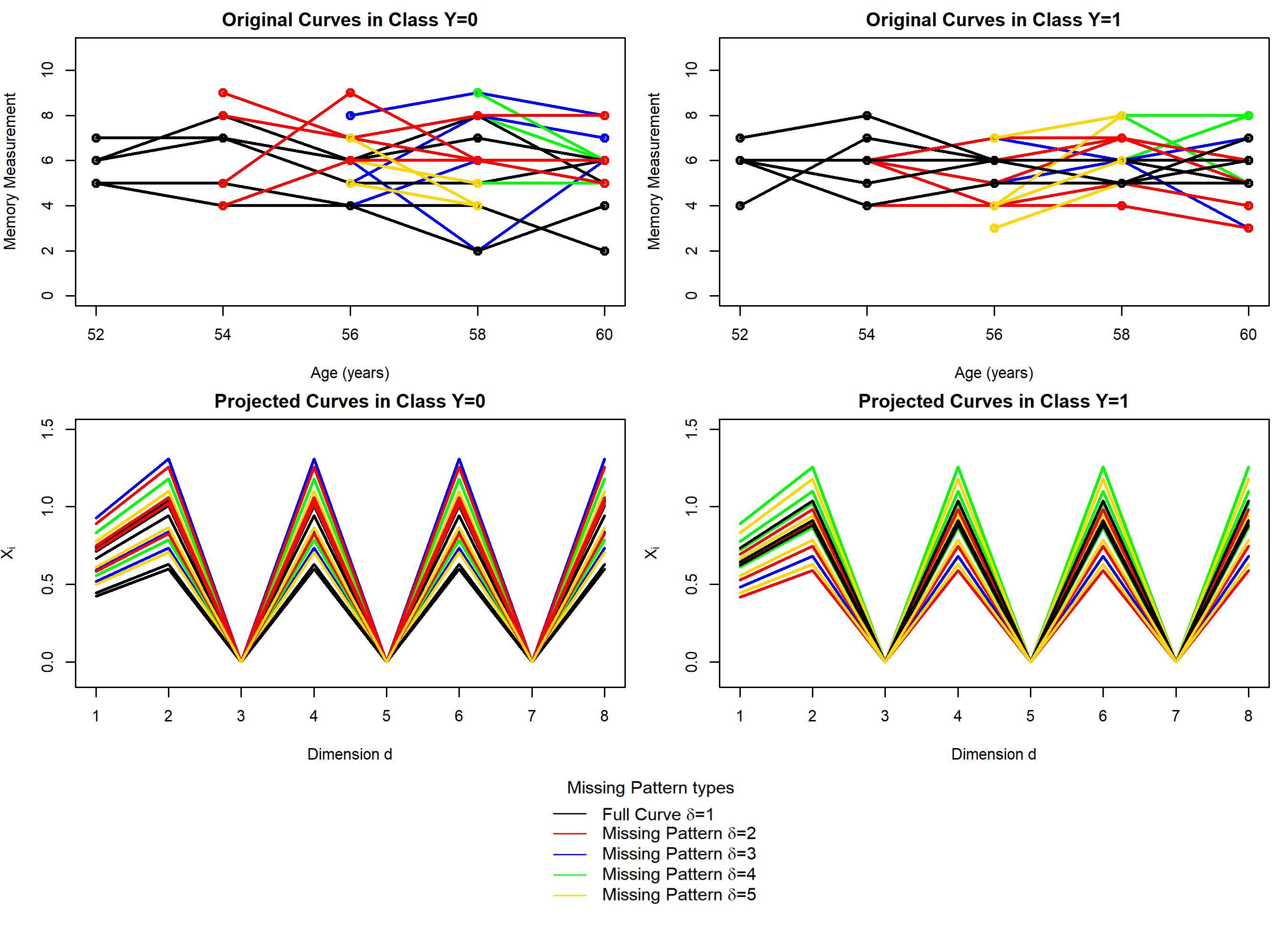} }}%
		\qquad
		\subfloat[]{{\includegraphics[scale = 0.33]{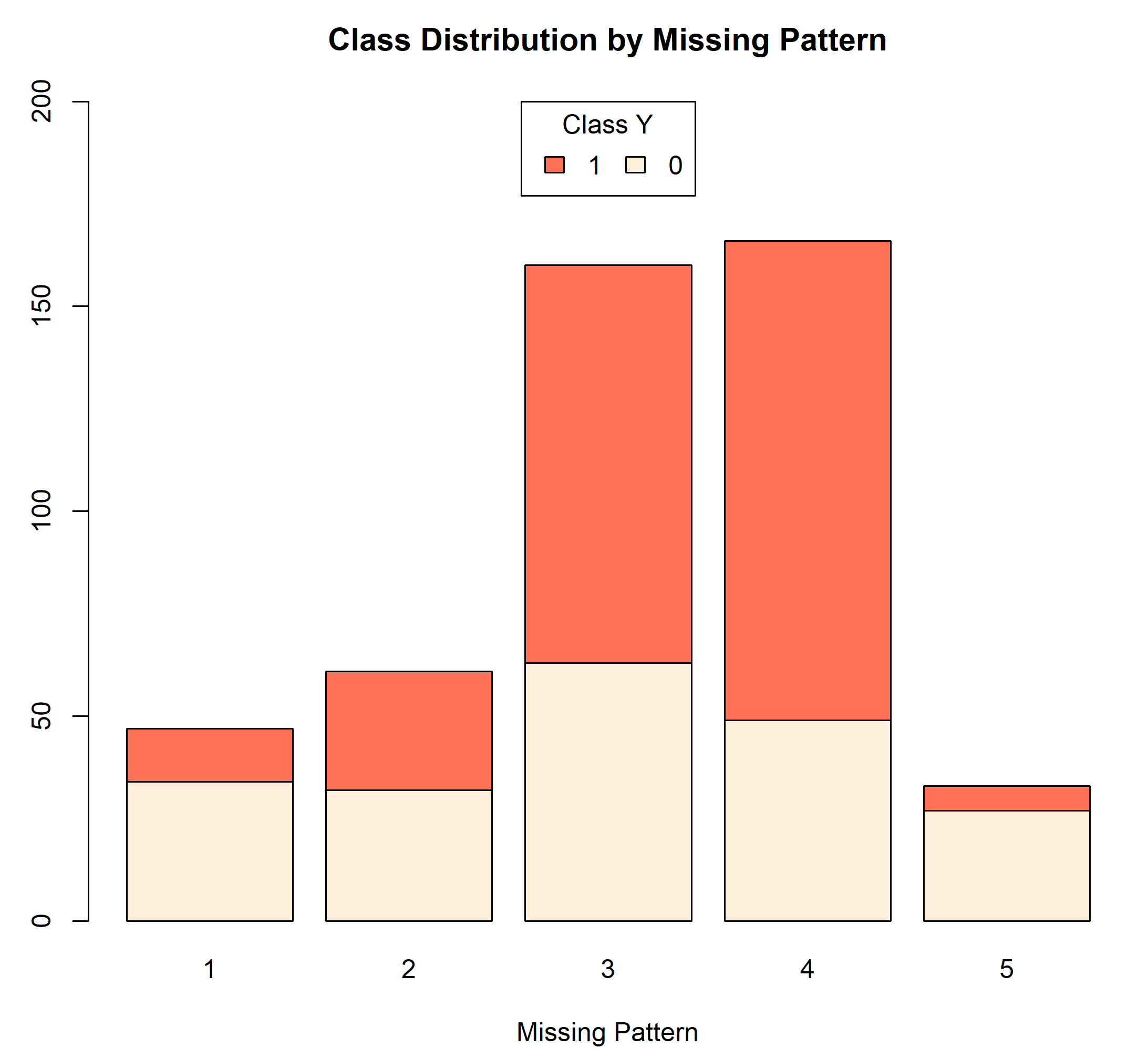} }}%
		\caption{\fontsize{10}{4}\selectfont (a) A sample of curves ${\boldsymbol{\mathcal{X}}}_i$ representing memory test scores measured at each visit for individuals in the Health and Retirement (HRS) study described in Example (C). The curves for those individuals whose data were available for every visit (complete) appear in solid black (in the top two plots) while those whose data were available for a subset of the visits (missing) are plotted using various colored lines according to their missing pattern.  The corresponding $d$-dimensional vector of the projected curves, $d = 1, \dots, 8$, are displayed below the original curves. (b) Distributions and proportion of class membership for each missing pattern in the utilized subset of the data.}%
		\label{memory_curves}%
	\end{figure}
	
	\vspace{4mm}\noindent
	{\it Example (C):  Health and Retirement Study.}
	
	\vspace{1mm}\noindent
	Advancements in medicine and health care have increased the life expectancy of individuals worldwide, resulting in the growth of the older population at an unprecedented rate.
	In 2015, the US Census Bureau reported that 8.5\% of people worldwide were aged 65 and over and projected that the number would increase to 17\% by 2050 (Wan et al. (2016)).  The Health and Retirement Study (HRS), conducted by the University of Michigan and funded by the National Institute on Aging and the Social Security Administration, is a longitudinal panel study aimed at helping researchers understand the challenges and opportunities related to an aging population.  The study began in 1992 with a representative cohort of 20,000 individuals aged 50 years and over and their spouses.  The study is ongoing with follow-up interviews conducted biennially. Data products related to the HRS can be found at {\tt https://hrs.isr.umich.edu/data-products}.\\
	
	The relationship between cognitive function and survival is widely studied; and it has been shown that cognitive decline is associated with survival into old age, see for example Eun and Choi (2017) and Mueller et al. (2017). Memory is assessed for partipants in the HRS using a short-term word-recall task.  At each interview, participants are given a list of 10 common nouns and asked to immediately recall as many words as they can from the list.  They are asked to recall the words again 5 minutes later.  The functional covariate ${\boldsymbol{ \mathcal{X}}}_i$ in this classification problem is the measured short-term memory score for participants at each visit, determined by their performance on the word-recall task.  Using memory score trajectories up to the age of 60, we study the survival of individuals past the age of 65.
	The class variable $y_i$ is coded as 0 = died by age 65 and 1 = survived past age 65.\\
	
	So that the number of missing patterns would be more tractable, the subset of the data used in this example contains 467 observations of individuals who were enrolled in the study at age 52.   Of these 467 observations, 90\% are fragmented curves. There are five distinct missing patterns observed in this subset of data, including the case where the memory scores are fully observable between the ages of 52 and 60.
	%Using the notation of section 2.1, we denote the observed missing patterns as $s_1 := \mathcal{I} = [52, 60], s_2 = [54, 60] \subset [52, 60], s_3 = [56, 60]  \subset [52, 60], s_4 = [58, 60] \subset [52, 60], s_5 = [56, 58] \subset [52, 60]$. 
	A sample of measured memory score trajectories and corresponding $d$-dimensional vectors of the projected curves is displayed in panel (a) of Figure \ref{memory_curves} while the distribution of each missing pattern is displayed in panel (b).\\
	
	To compare the performance of our proposed classifier, $\widehat\Gamma_n$, with that of the classifier based on the complete case analysis, $\widehat \Gamma_{\scriptscriptstyle CC}$ 
	%and the classifier based on the entire fragmented curve, $\widehat \Gamma_{\scriptscriptstyle F}$, 
	the sample of $n = 467$ individuals was split into a training sequence of size $n' = 233$ and a testing sequence of size $n - n' = 234$. Table \ref{memory_table} provides the average error rates of each classifier committed on the testing sequence over 200 such sample splits with standard errors given in parenthesis as well as a visual display of classifier performance.  One notices that the advantage of the proposed classifier $\widehat\Gamma_n$ over the complete case classifier seems less dramatic than expected, given the extremely large proportion of fragmented curves.  This illustrates that the success of the classifier is also a function of the correlation between the outcome and the functional covariate.  Though cognitive function (measured through memory) is certainly correlated with survival, it is not the strongest single predictor of survival.\\ 
	
	\begin{table}[H]
		\captionsetup{singlelinecheck = false, justification=justified,font = footnotesize}
		\centering
		\begin{footnotesize}
			\caption{Error rates for $\widehat \Gamma_n$ (the classifier based on filtered curves) and 
				%$\widehat \Gamma_{\scriptscriptstyle F}$ (the classifier based on the available fragmented curve) and 
				$\widehat \Gamma_{\scriptscriptstyle CC}$ (the complete case analysis).}
			\begin{tabular}{ccccc}
				& & & \hspace{50pt} & \multirow{1}{*}{\includegraphics[scale = 0.375]{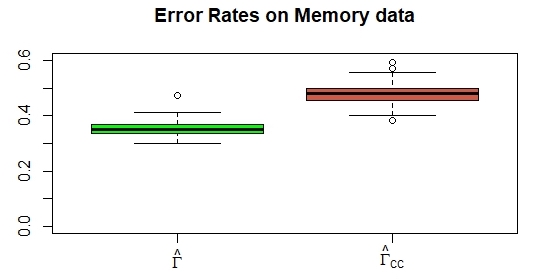}}\\
				\cmidrule[1.25pt]{1-3}
				$\%$ Missing Data & $\widehat \Gamma_n$ & $\widehat \Gamma_{\scriptscriptstyle CC}$ & \hspace{50pt} &\\
				\cmidrule[1.25pt]{1-3}
				$90\%$ & 0.3502 & $0.4764$ & \hspace{50pt} &\\
				& (0.0218) & $(0.0303)$ & \hspace{50pt} &\\
				\cmidrule[1.25pt]{1-3}
				\label{memory_table}
			\end{tabular}
		\end{footnotesize}
	\end{table}

%%%%%%%%%%%%%%%%%%%%%%%%%%%%%%%%%%%%%%%%%%%%%%%%%%%%%%%%%%%%%%%%%%%%%%%%%%%%%%%%
%%%%%%%%%%%%%%%%%%%%%%%%%%%%%%%%%%%%%%%%%%%%%%%%%%%%%%%%%%%%%%%%%%%%%%%%%%%%%%%%
%%%%%%%%%%%%%%%%%%%%%%%   Proofs Begin Here:  %%%%%%%%%%%%%%%%%%%%%%%%%%%%%%%%%%
%%%%%%%%%%%%%%%%%%%%%%%%%%%%%%%%%%%%%%%%%%%%%%%%%%%%%%%%%%%%%%%%%%%%%%%%%%%%%%%%
%%%%%%%%%%%%%%%%%%%%%%%%%%%%%%%%%%%%%%%%%%%%%%%%%%%%%%%%%%%%%%%%%%%%%%%%%%%%%%%%

\newpage
\section{Proofs}\label{proof}
PROOF OF THEOREM \ref{B1}\\
The proof of Part (i) is virtually the same as that of Theorem \ref{A}, whereas the proof of Part (iii) is exactly the same as that of Lemma 1 of Mojirsheibani and Shaw (2018). 

\vspace{3mm}
\noindent
The proof of Part (ii):\\
Let $\{\psi_1, \psi_2, \dots\}$ be a complete orthonormal basis in $L^2(s)$, $s\subset \mathbb{R}$, and for any $\boldsymbol{\chi}\in L^2(s)$ define the map $t_s:L^2(s)\to\ell_2$ by $t_s(\boldsymbol{\chi})= (\langle \boldsymbol{\chi}, \psi_1\rangle_s\,,\, \langle \boldsymbol{\chi}, \psi_2\rangle_s\,, \dots)$. Since 
\begin{equation} \label{NEW2}
{\bf X}^{(k)} = t_k(\boldsymbol{\chi}|_{s_k}) :=(\langle \boldsymbol{\chi}, \psi_1\rangle_{s_k}\,,\, \langle \boldsymbol{\chi}, \psi_2\rangle_{s_k}\,, \dots),
\end{equation} 
the function $\phi_k({\bf X}^{(k)})$ in (\ref{E8}) can equivalently be written as
{\allowdisplaybreaks
\begin{eqnarray}
&&\phi_k({\bf X}^{(k)}) = \phi_k(t_k(\boldsymbol{\chi}|_{s_k})) \nonumber\\
&& ~~= \mathbb{E}\left[\mathbb{E}\left\{Y I\{\delta=k\} - (1-Y)I\{\delta=k\} \Big| t_k(\boldsymbol{\chi}|_{s_k}), Y\right\}\Big|t_k(\boldsymbol{\chi}|_{s_k})
\right]\nonumber\\
&& ~~= \mathbb{E}\Big[Y\, \mathbb{P}\left\{\delta= k\big|t_k(\boldsymbol{\chi}|_{s_k}), Y\right\}\Big|t_k(\boldsymbol{\chi}|_{s_k})\Big]-
\mathbb{E}\Big[(1-Y)\, \mathbb{P}\left\{\delta= k\big|t_k(\boldsymbol{\chi}|_{s_k}), Y\right\}\Big|t_k(\boldsymbol{\chi}|_{s_k})\Big]~~~~~
%\nonumber\\
\label{PHI2}
\end{eqnarray}
Also, observe that with $\Gamma^{\mbox{\tiny B}}$ as in (\ref{E9}), 
\begin{eqnarray*}
&&	\mathbb{P}\left\{\Gamma^{\mbox{\tiny B}}({\bf X}^{(\delta)}) =Y \right\}\\
&&= \mathbb{P}\left\{\Gamma^{\mbox{\tiny B}}({\bf X}^{(\delta)}) =1, Y=1 \right\}+ \mathbb{P}\left\{\Gamma^{\mbox{\tiny B}}({\bf X}^{(\delta)}) =0, Y=0 \right\}\\
&&= \sum_{k=1}^M\Big[\mathbb{P}\big\{\delta=k, Y=1, \phi_k(t_k(\boldsymbol{\chi}|_{s_k}))>0 \big\}+ \mathbb{P}\big\{\delta=k, Y=0, \phi_k(t_k(\boldsymbol{\chi}|_{s_k}))\leq 0 \big\}\Big] \\
&&= \sum_{k=1}^M \mathbb{E} \left[Y I\left\{\phi_k(t_k(\boldsymbol{\chi}|_{s_k}))>0 \right\}\cdot \mathbb{P}\left\{\delta=k \Big|t_k(\boldsymbol{\chi}|_{s_k}), Y \right\}\right]\\
&& ~~~~~~~~~~~~~~~~~~~~ +~\sum_{k=1}^M \mathbb{E} \left[(1-Y) I\left\{\phi_k(t_k(\boldsymbol{\chi}|_{s_k}))\leq 0 \right\}\cdot \mathbb{P}\left\{\delta=k \Big|t_k(\boldsymbol{\chi}|_{s_k}), Y \right\}\right]\\
&&= \sum_{k=1}^M \mathbb{E} \left[ I\Big\{ \phi_k(t_k(\boldsymbol{\chi}|_{s_k}))>0 \Big\}\cdot \mathbb{E}\left(Y\cdot \mathbb{P}\left\{\delta=k \Big|t_k(\boldsymbol{\chi}|_{s_k}), Y \right\}\bigg|t_k(\boldsymbol{\chi}|_{s_k})\right)\right]\\
&& ~~~~~~~~~~~~~~~~~~~~ +~\sum_{k=1}^M \mathbb{E} \left[ I\Big\{ \phi_k(t_k(\boldsymbol{\chi}|_{s_k}))\leq 0 \Big\}\cdot \mathbb{E}\left((1-Y)\cdot \mathbb{P}\left\{\delta=k \Big|t_k(\boldsymbol{\chi}|_{s_k}), Y \right\}\bigg|t_k(\boldsymbol{\chi}|_{s_k})\right)\right]\\
&&= \sum_{k=1}^M \mathbb{E} \Bigg[\max \Bigg\{
\mathbb{E}\left(Y\cdot \mathbb{P}\left\{\delta=k \Big|t_k(\boldsymbol{\chi}|_{s_k}), Y \right\}\bigg|t_k(\boldsymbol{\chi}|_{s_k})\right)\,,\\
&& ~~~~~~~~~~~~~~~~~~~~~~~~
\mathbb{E}\left((1-Y)\cdot \mathbb{P}\left\{\delta=k \Big|t_k(\boldsymbol{\chi}|_{s_k}), Y \right\} \bigg|t_k(\boldsymbol{\chi}|_{s_k})\right)
\Bigg\}\Bigg],
\end{eqnarray*}
where the last equality above follows from the representation of $\phi_k(t_k(\boldsymbol{\chi}|_{s_k}))$ in (\ref{PHI2}). However, using the fact that  $\mathbb{P}\left\{\delta=k \big|t_k(\boldsymbol{\chi}|_{s_k}), Y \right\} =\mathbb{E}\left[ \mathbb{P}\left\{\delta=k \big|\,\boldsymbol{\chi}|_{s_k}, Y \right\}\big| t_k(\boldsymbol{\chi}|_{s_k}), Y  \right]$, we can write 
\begin{eqnarray*}
	&&	\mathbb{P}\left\{\Gamma^{\mbox{\tiny B}}({\bf X}^{(\delta)}) =Y \right\}\\
	&&= \sum_{k=1}^M \mathbb{E} \Bigg[\max \Bigg\{
	\mathbb{E}\left(\mathbb{E}\Big[Y\cdot \mathbb{P}\left\{\delta=k \Big|\boldsymbol{\chi}|_{s_k}, Y \right\}\Big|t_k(\boldsymbol{\chi}|_{s_k}), Y\Big]\bigg| t_k(\boldsymbol{\chi}|_{s_k}) \right)\,,\\
	&& ~~~~~~~~~~~~~~~~~~~~~~~~
	\mathbb{E}\left(\mathbb{E}\Big[(1-Y)\cdot \mathbb{P}\left\{\delta=k \Big|\boldsymbol{\chi}|_{s_k}, Y \right\}\Big|t_k(\boldsymbol{\chi}|_{s_k}), Y\Big]\bigg| t_k(\boldsymbol{\chi}|_{s_k}) \right)\Bigg\}\Bigg]\\
	&&= \sum_{k=1}^M \mathbb{E} \Bigg[\max \Bigg\{
	\mathbb{E}\left(Y \mathbb{P}\left\{\delta=k \Big|\boldsymbol{\chi}|_{s_k}, Y\right\}\bigg|t_k(\boldsymbol{\chi}|_{s_k})\right)\,,~
	\mathbb{E}\left((1-Y) \mathbb{P}\left\{\delta=k \Big|\boldsymbol{\chi}|_{s_k}, Y\right\}\bigg|t_k(\boldsymbol{\chi}|_{s_k})\right)
	\Bigg\}\Bigg]\\
	&&= \sum_{k=1}^M \mathbb{E} \Bigg[\max \Bigg\{
	\mathbb{E}\left[\mathbb{E}\left(Y \mathbb{P}\left\{\delta=k \Big|\boldsymbol{\chi}|_{s_k}, Y\right\} \bigg|\boldsymbol{\chi}|_{s_k}\right) \bigg|t_k(\boldsymbol{\chi}|_{s_k})\right]\,,\\
	&& ~~~~~~~~~~~~~~~~~~~~~~~~
	\mathbb{E}\left[\mathbb{E}\left((1-Y) \mathbb{P}\left\{\delta=k \Big|\boldsymbol{\chi}|_{s_k}, Y\right\} \bigg|\boldsymbol{\chi}|_{s_k}\right) \bigg|t_k(\boldsymbol{\chi}|_{s_k})\right]\Bigg\}\Bigg]\\
	&&\leq\, 	\sum_{k=1}^M \mathbb{E} \Bigg[\mathbb{E} \Bigg( \max \Bigg\{ \mathbb{E}\left(Y \mathbb{P}\left\{\delta=k \Big|\boldsymbol{\chi}|_{s_k}, Y\right\} \bigg|\boldsymbol{\chi}|_{s_k}\right)\, ,\\
	&& ~~~~~~~~~~~~~~~~~~~
	\mathbb{E}\left((1-Y)\, \mathbb{P}\left\{\delta=k \Big|\boldsymbol{\chi}|_{s_k}, Y\right\} \bigg|\boldsymbol{\chi}|_{s_k}\right)\Bigg\}\Bigg| t_k(\boldsymbol{\chi}|_{s_k})\Bigg)\Bigg],~\mbox{(by Jensen's inequality)}\\
	%&& ~~~~~~~~~~\mbox{(Jensen's inequality)}\\
	&&= \sum_{k=1}^M \mathbb{E} \Bigg[\max \Bigg\{\mathbb{E}\left(Y \mathbb{P}\left\{\delta=k \Big|\boldsymbol{\chi}|_{s_k}, Y\right\} \bigg|\boldsymbol{\chi}|_{s_k}\right),\,
	\mathbb{E}\left((1-Y) \mathbb{P}\left\{\delta=k \Big|\boldsymbol{\chi}|_{s_k}, Y\right\} \bigg|\boldsymbol{\chi}|_{s_k}\right)\Bigg\}\Bigg]\\
	&&= \mathbb{P}\left\{\Gamma_{0}^{\mbox{\tiny B}} (\boldsymbol{\chi}^{(\delta)}) =Y \right\},
\end{eqnarray*}
i.e., $\mathbb{P}\left\{\Gamma^{\mbox{\tiny B}}({\bf X}^{(\delta)}) \neq Y \right\} \geq\, \mathbb{P}\left\{\Gamma_{0}^{\mbox{\tiny B}} (\boldsymbol{\chi}^{(\delta)}) \neq Y \right\}$,
where $\Gamma_{0}^{\mbox{\tiny B}} (\boldsymbol{\chi}^{(\delta)})$ is the optimal classifier defined in (\ref{E4}); see Theorem \ref{A}.  To complete the proof, it is sufficient to show that the map $t_k$ in (\ref{NEW2}) is one-to-one (and thus invertible). But, observe that for ${\bf u}, {\bf v} \in L^2(s_k)$, we have $t_k({\bf u}-{\bf v}) = 0$ if and only if ${\bf u}-{\bf v}=0$ (by the completeness of the basis $\{\psi_1, \psi_2, \dots\}$); the result now follows since $t_k({\bf u})-t_k({\bf v}) =t_k({\bf u}-{\bf v})$.

\hfill $\Box$
}  % this "}" corresponds to {\allowdisplaybreaks above.

\vspace{3mm}
\noindent
PROOF OF THEOREM \ref{B2}\\
We recall that if $\delta=k$ then ${\bf X}^{(d,k)}$ is the observed $d$-dim covariate, where $k=1,\dots,M$. This means that when $\delta=k$, a classifier is any function of the form $g_{d,k}: \mathbb{R}^d \to \{0,1\}$. Therefore the general classifier, denoted by $\Gamma^{d}$, can always be written in the form
\begin{equation}\label{E9D}
\Gamma^{d}({\bf X}^{(d,\delta)}) := \sum_{k=1}^M I\{\delta=k\}\cdot g_{d,k}({\bf X}^{(d,k)}).
\end{equation}
Now, for $k=1,\dots,M$, define the functions
%\begin{eqnarray*}
$	r_k({\bf x},y) = \mathbb{P}\left\{\delta=k\,\big|\,{\bf X}^{(d,k)}={\bf x}, Y=y\right\}, ~~y=0,1,$ and $\eta_k({\bf x}) = \mathbb{P}\left\{Y=1\,\big|\,{\bf X}^{(d,k)}={\bf x}\right\}\, =\, \mathbb{E}\left[Y\big|{\bf X}^{(d,k)}={\bf x}\right],$
%\end{eqnarray*}
and observe that the function $\phi_{d,k}$ in (\ref{E8B}) can be written as
\begin{eqnarray}
\phi_{d,k}({\bf X}^{(d,k)}) &=& \mathbb{E}\left\{\mathbb{E}\left[(2Y-1) I\{\delta=k\}\,\Big|\, {\bf X}^{(d,k)},\, Y\right]\Big|{\bf X}^{(d,k)}\right\},\nonumber\\
&=& \mathbb{E}\left[(2Y-1) \mathbb{P}\{\delta=k\big|{\bf X}^{(d,k)},\, Y\}\,\Big|\, {\bf X}^{(d,k)}\right]\nonumber\\
&=& \mathbb{E}\left[(2Y-1) \Big(Y\cdot r_k({\bf X}^{(d,k)},1)+ (1-Y)r_k({\bf X}^{(d,k)},0)  \Big)\,\Big|\, {\bf X}^{(d,k)}\right]\nonumber\\
&=& \mathbb{E}\left[Y\cdot r_k({\bf X}^{(d,k)},1)+ (Y-1)\cdot r_k({\bf X}^{(d,k)},0)  \,\Big|\, {\bf X}^{(d,k)}\right]\,,~ \mbox{(because $Y^2$=$Y$)}\nonumber\\
&=& \eta_k({\bf X}^{(d,k)})r_k({\bf X}^{(d,k)},1) + \Big(\eta_k({\bf X}^{(d,k)})-1\Big) r_k({\bf X}^{(d,k)},0).  \label{FK1}
\end{eqnarray}
Therefore, the classifier $\Gamma^{\mbox{\tiny B},d}$ in (\ref{E9B}) can be written as
\[
\Gamma^{\mbox{\tiny B},d}({\bf X}^{(d,\delta)}) = \sum_{k=1}^M I\{\delta=k\}\cdot I\left\{\eta_k({\bf X}^{(d,k)})r_k({\bf X}^{(d,k)},1) + \Big(\eta_k({\bf X}^{(d,k)})-1\Big) r_k({\bf X}^{(d,k)},0) > 0\right\},
\]
and this can be used to write
\begin{eqnarray*}
	&&\mathbb{P}\left\{\Gamma^{\mbox{\tiny B},d}({\bf X}^{(d,\delta)}) =Y\right\} \\
	% && ~~= \mathbb{P}\left\{\Gamma^{\mbox{\tiny B},d}({\bf X}^{(d,\delta)}) =1, Y=1 \right\}+ \mathbb{P}\left\{\Gamma^{\mbox{\tiny B},d}({\bf X}^{(d,\delta)}) =0, Y=0 \right\}\\
	&& ~~=\sum_{k=1}^M \mathbb{P}\left\{Y=1, \delta=k, [\phi_{d,k}({\bf X}^{(d,k)})>0]\right\} + \sum_{k=1}^M \mathbb{P}\left\{Y=0, \delta=k, [\phi_{d,k}({\bf X}^{(d,k)})\leq 0]\right\}\\
	&& ~~:=\sum_{k=1}^M\pi_{k1} + \sum_{k=1}^M\pi_{k0}\,,~~ (\mbox{say}).
\end{eqnarray*}
But 
\begin{eqnarray*}
	\pi_{k1} &=& \mathbb{E}\Big[I\{Y=1\}\cdot I\left\{\phi_{d,k}({\bf X}^{(d,k)})>0 \right\}\cdot \mathbb{P}\left\{\delta =k\big|{\bf X}^{(d,k)}, Y\right\}\Big]\\
	&=& \mathbb{E}\Big[I\left\{\phi_{d,k}({\bf X}^{(d,k)})>0 \right\}\cdot r_k({\bf X}^{(d,k)},1)\cdot \mathbb{E}\big[I\{Y=1\}\big|{\bf X}^{(d,k)}\big]\Big]\\
	&=& \mathbb{E}\Big[I\left\{\phi_{d,k}({\bf X}^{(d,k)})>0 \right\}\cdot r_k({\bf X}^{(d,k)},1)\cdot \eta_k({\bf X}^{(d,k)})\Big].
\end{eqnarray*}
Also, similar arguments yield
$
\pi_{k0} = \mathbb{E}\big[I\left\{\phi_{d,k}({\bf X}^{(d,k)})\leq 0 \right\}\cdot r_k({\bf X}^{(d,k)},0)\cdot \big(1-\eta_k({\bf X}^{(d,k)})\big)\big].
$
Thus, we have
\begin{eqnarray*}
	\mathbb{P}\{\Gamma^{\mbox{\tiny B},d}({\bf X}^{(d,\delta)})= Y\} &=&
	\sum_{k=1}^M \Bigg(\mathbb{E}\Big[I\left\{\phi_{d,k}({\bf X}^{(d,k)})>0 \right\}\cdot r_k({\bf X}^{(d,k)},1)\cdot \eta_k({\bf X}^{(d,k)})\Big]\\
	&& ~~~+ \mathbb{E}\Big[I\left\{\phi_{d,k}({\bf X}^{(d,k)})\leq 0 \right\}\cdot r_k({\bf X}^{(d,k)},0)\cdot \Big(1-\eta_k({\bf X}^{(d,k)})\Big)\Big]\Bigg).
\end{eqnarray*}
Furthermore, for any other classifier $\Gamma^{d}({\bf X}^{(d,\delta)})$ given by (\ref{E9D}), it is not difficult to see that
\begin{eqnarray*}
	\mathbb{P}\{\Gamma^{d}({\bf X}^{(d,\delta)})= Y\} &=&
	\sum_{k=1}^M \Bigg(\mathbb{E}\Big[I\left\{g_{d,k}({\bf X}^{(d,k)})=1 \right\}\cdot r_k({\bf X}^{(d,k)},1)\cdot \eta_k({\bf X}^{(d,k)})\Big]\\
	&& ~~~+ \mathbb{E}\Big[I\left\{g_{d,k}({\bf X}^{(d,k)})=0 \right\}\cdot r_k({\bf X}^{(d,k)},0)\cdot \Big(1-\eta_k({\bf X}^{(d,k)})\Big)\Big]\Bigg).
\end{eqnarray*}
Therefore,
\begin{eqnarray}
	&& \mathbb{P}\{\Gamma^{\mbox{\tiny B},d}({\bf X}^{(d,\delta)})\neq Y\} -
	\mathbb{P}\{\Gamma^{d}({\bf X}^{(d,\delta)})\neq Y\}\nonumber\\
	&& = \sum_{k=1}^M\mathbb{E}\Big[\Big(I\left\{\phi_{d,k}({\bf X}^{(d,k)})>0 \right\}-I\left\{g_{d,k}({\bf X}^{(d,k)})=1 \right\}\Big)\cdot r_k({\bf X}^{(d,k)},1)\cdot \eta_k({\bf X}^{(d,k)})\Big]\nonumber\\
	&& ~~ + \sum_{k=1}^M\mathbb{E}\Big[\Big(I\left\{\phi_{d,k}({\bf X}^{(d,k)})\leq 0 \right\}-I\left\{g_{d,k}({\bf X}^{(d,k)})=0 \right\}\Big)\cdot r_k({\bf X}^{(d,k)},0)\cdot \big(1-\eta_k({\bf X}^{(d,k)})\big)\Big]~~\nonumber\\
	&& = \sum_{k=1}^M\mathbb{E}\Bigg[I\bigg\{g_{d,k}({\bf X}^{(d,k)}) \neq
	I\left\{\phi_{d,k}({\bf X}^{(d,k)})>0 \right\} \bigg\}\label{NEW8}\\
	&& ~~~~~~~~~~~~~~~~~~\times \bigg| r_k({\bf X}^{(d,k)},1)\cdot \eta_k({\bf X}^{(d,k)}) - r_k({\bf X}^{(d,k)},0)\cdot \big(1-\eta_k({\bf X}^{(d,k)})\big)\bigg|\Bigg]\nonumber\\
	&& \geq 0, \nonumber
\end{eqnarray}
where (\ref{NEW8}) follows from the definitions of $\Gamma^{\mbox{\tiny B},d}$ and $\Gamma^{d}$ in conjunction with the expression in (\ref{FK1}). 

\hfill $\Box$\\

\vspace{1mm}\noindent
In order to prove Theorem \ref{C}, we first state a number of lemmas. In what follows, we use the following notation: 
\begin{eqnarray}
\mathcal{R}_m(d,h_1,\dots,h_M) &=& \mathbb{P}\left\{\Gamma_m^{d}({\bf X}^{(d,\delta)})\neq Y \big| \mathbb{D}_m\right\} \label{R2}\\
\widehat{\mathcal{R}}_{m,\ell}(d,h_1,\dots,h_M) &=&	\ell^{-1} \sum_{i:~ {(\bf X}^{(\delta_i)}, Y_i,\delta_i) \in \mathbb{D}_{\ell}} I\big\{\Omega_i(m,d,h_1,\dots, h_M)\big\},  \label{R3}
\end{eqnarray}
where $\Gamma_m^{d}({\bf X}^{(d,\delta)})$ and $\Omega_i(m,d,h_1,\dots, h_M)$ are as in (\ref{E11}) and (\ref{E13}), respectively.
\begin{lem}\label{LEM-1}
	Let $\widehat{\mathcal{R}}_{m,\ell}$ and $\mathcal{R}_m$ be as in (\ref{R3}) and (\ref{R2}). If $\ell^{-1} \log |\mathcal{H}_n| \to 0$ and $\ell^{-1} \log d_n \to 0$, where $|\mathcal{H}_n|$ is the cardinality of the set $\mathcal{H}_n$,
	then, as $n\to\infty$,
	\begin{eqnarray*}
	 \sup_{1\leq d \leq d_n\,,\, h_1,\dots,h_M\in \mathcal{H}_n~} 
		\left|\widehat{\mathcal{R}}_{m,\ell}(d,h_1,\dots,h_M)- \mathcal{R}_m(d,h_1,\dots,h_M)\right| &\longrightarrow^{\mbox{a.s.}}& 0\,.
	\end{eqnarray*}
\end{lem}

\vspace{2mm}\noindent
PROOF OF LEMMA \ref{LEM-1}\\
First observe that for any given constant $\beta>0$,
\begin{eqnarray}
&& \mathbb{P}\left\{\sup_{1\leq d \leq d_n\,,\, h_1,\dots,h_M\in \mathcal{\mathcal{H}}_n~} 
\left|\widehat{\mathcal{R}}_{m,\ell}(d,h_1,\dots,h_M)- \mathcal{R}_m(d,h_1,\dots,h_M)\right| > \beta \right\}  \nonumber\\
&& ~\leq ~\sum_{1\leq d\leq d_n} \sum_{~h_1,\dots,h_M\in \mathcal{H}_n}
\mathbb{P}\left\{ 
\left|\widehat{\mathcal{R}}_{m,\ell}(d,h_1,\dots,h_M)- \mathcal{R}_m(d,h_1,\dots,h_M)\right| > \beta \right\}  \nonumber\\
&& ~\leq~ d_n |\mathcal{H}_n|^M  \sup_{1\leq d\leq d_n}\sup_{h_1,\dots,h_M\in \mathcal{H}_n}
\mathbb{E}\left[\mathbb{P}\left\{\left|\widehat{\mathcal{R}}_{m,\ell}(d,h_1,\dots,h_M)- \mathcal{R}_m(d,h_1,\dots,h_M)\right| > \beta\bigg| \mathbb{D}_m\right\}\right]~~~\nonumber
\end{eqnarray}
where $|\mathcal{H}_n|$ is the cardinality of the set $\mathcal{H}_n$.
But, with $\Omega_i(m,d,h_1,\dots, h_M)$ as in (\ref{E13}),
\begin{eqnarray*}
&& \mathbb{P}\left\{\left|\widehat{\mathcal{R}}_{m,\ell}(d,h_1,\dots,h_M)- \mathcal{R}_m(d,h_1,\dots,h_M)\right| > \beta\,\Big|\, \mathbb{D}_m\right\}\\
&& ~~=~ \mathbb{P}\Bigg\{\bigg|
\ell^{-1} \sum_{i:~ {(\bf X}^{(\delta_i)}, Y_i,\delta_i) \in \mathbb{D}_{\ell}} I\big\{\Omega_i(m,d,h_1,\dots, h_M)\big\} - \mathbb{P}\big\{\Omega_1(m,d,h_1,\dots, h_M)\big\}\bigg|> \beta\,\Bigg|\,\mathbb{D}_m\Bigg\}~~~~\\
&& ~~\leq~ 2\, e^{-2\ell\beta^2},~~~~~~\mbox{(by Hoeffding's inequality),}
\end{eqnarray*}
which does not depend on $\mathbb{D}_m$ or any of the parameters $d, h_1,\dots,h_M$. Therefore
\begin{eqnarray}
\mathbb{P}\left\{\sup_{1\leq d \leq d_n\,,\, h_1,\dots,h_M\in \mathcal{H}_n~} 
\left|\widehat{\mathcal{R}}_{m,\ell}(d,h_1,\dots,h_M)- \mathcal{R}_m(d,h_1,\dots,h_M)\right| > \beta \right\} 
&\leq& 2 \,d_n |\mathcal{H}_n|^M e^{-2\ell\beta^2}. \nonumber
\end{eqnarray}
Furthermore, the conditions of Lemma \ref{LEM-1} imply that $\sum_{n=1}^{\infty} d_n |\mathcal{H}_n|^M e^{-\ell\beta^2/2} < \infty$. The result  now follows from an application of the Borel-Cantelli lemma.

\hfill $\Box$

\begin{lem}\label{LEM-2}
	Let $\widehat{\Gamma}_n({\bf X}^{(\widehat{d},\delta)})$ be the classifier in (\ref{E14}). Also, let $\widehat{\mathcal{R}}_{m,\ell}$ and $\mathcal{R}_m$ be as in (\ref{R3}) and (\ref{R2}). Then
	\begin{eqnarray*}
		&& \mathbb{P}\left\{\widehat{\Gamma}_n({\bf X}^{(\widehat{d},\delta)}) \neq Y \Big| \mathbb{D}_n\right\}~ - \inf_{1\leq d \leq d_n\,,\, h_1,\dots,h_M\in \mathcal{H}_n~} \mathcal{R}_m(d,h_1,\dots,h_M)\\
		&&~~~~~~\leq ~~ 2~ \sup_{1\leq d \leq d_n\,,\, h_1,\dots,h_M\in \mathcal{H}_n~} 
		\left|\widehat{\mathcal{R}}_{m,\ell}(d,h_1,\dots,h_M)- \mathcal{R}_m(d,h_1,\dots,h_M)\right|.
	\end{eqnarray*}
\end{lem}
PROOF of LEMMA \ref{LEM-2}\\
The proof of this lemma, which is similar to that of Lemma 8.2 of Devroye et al (1996), is straightforward and will not be given here.
\hfill $\Box$ 

\begin{lem}\label{LEM-3}
	Let $\Gamma^{\mbox{\tiny B},d}({\bf X}^{(d,\delta)})$ be the classifier defined via (\ref{E9B}) and (\ref{E8B}). Also, let $g_{d,k}$ in the definition of the classifier $\Gamma^{d}({\bf X}^{(d,\delta)})$ in (\ref{E9D}) be of the form $g_{d,k}({\bf X}^{(d,k)})= I\{G_{d,k}({\bf X}^{(d,k)}) > 0\} $ for some function $G_{d,k} : \mathbb{R}^d \to [-1, 1]$. Then  
	\begin{eqnarray*}
	\mathbb{P}\left\{\Gamma^{d}({\bf X}^{(d,\delta)})\neq Y\right\}-
	\mathbb{P}\left\{\Gamma^{\mbox{\tiny B},d}({\bf X}^{(d,\delta)})\neq Y\right\} &\leq& \sum_{k=1}^M \mathbb{E}\Big|\phi_{d,k}({\bf X}^{(d,k)})-G_{d,k}({\bf X}^{(d,k)}) \Big|\,,
	\end{eqnarray*}
	where $\Gamma^{\mbox{\tiny B},d}({\bf X}^{(d,\delta)})$ and $\phi_{d,k}({\bf X}^{(d,k)})$ are as in (\ref{E9B}) and (\ref{E8B}), respectively.
\end{lem}
PROOF OF LEMMA \ref{LEM-3}\\
The expression in (\ref{NEW8}) in the proof of Theorem \ref{B2} shows that in view of (\ref{FK1}) one has 
\begin{eqnarray*}
	&& \mathbb{P}\left\{\Gamma^{d}({\bf X}^{(d,\delta)})\neq Y\right\}-
	\mathbb{P}\left\{\Gamma^{\mbox{\tiny B},d}({\bf X}^{(d,\delta)})\neq Y\right\} \\
	&& ~~~~~\leq~ \sum_{k=1}^M \mathbb{E}\bigg(
	I\Big\{g_{d,k}({\bf X}^{(d,k)}) \neq I\left\{\phi_{d,k}({\bf X}^{(d,k)})>0\right\}\Big\}\times \Big|\phi_{d,k}({\bf X}^{(d,k)})\Big| \bigg)\,.
\end{eqnarray*}
Given the definition of $g_{d,k}({\bf X}^{(d,k)})$ in the statement of the lemma, it is straightforward to see that on the set $\big\{g_{d,k}({\bf X}^{(d,k)}) \neq I\left\{\phi_{d,k}({\bf X}^{(d,k)})>0\right\}\big\}$, one has 
\begin{eqnarray*}
	\mathbb{E}\bigg(
	I\Big\{g_{d,k}({\bf X}^{(d,k)}) \neq I\left\{\phi_{d,k}({\bf X}^{(d,k)})>0\right\}\Big\}\times \Big|\phi_{d,k}({\bf X}^{(d,k)})\Big| \bigg) 
	&\leq& \mathbb{E}\Big|\phi_{d,k}({\bf X}^{(d,k)})-G_{d,k}({\bf X}^{(d,k)})\Big|,
\end{eqnarray*}
which completes the proof of the lemma.

\hfill $\Box$

\vspace{2mm}\noindent
The following result is an immediate corollary to Lemma \ref{LEM-3}.
\begin{cor} \label{COR-1}
	Let $\Gamma^{\mbox{\tiny B},d}({\bf X}^{(d,\delta)})$ be the classifier defined via (\ref{E9B}) and (\ref{E8B}). Also, for $k=1,\dots, M$, let $G_{m,d,k}({\bf X}^{(d,k)})$ be any sample-based version of the function $G_{d,k}({\bf X}^{(d,\delta)})$ that appears in Lemma \ref{LEM-3}, based on the training sample $\mathbb{D}_m$, and consider the classifier
	\[
		\widetilde{\Gamma}_m({\bf X}^{(d,\delta)})=\sum_{k=1}^M I\{\delta=k\}\cdot g_{m,d,k}({\bf X}^{(d,k)}),
	\]
	where $g_{m,d,k}({\bf X}^{(d,k)}) = I\left\{G_{m,d,k}({\bf X}^{(d,k)}) > 0\right\}$. Then
	\begin{eqnarray*}
	\mathbb{P}\left\{\widetilde{\Gamma}_m({\bf X}^{(d,\delta)})\neq Y \Big| \mathbb{D}_m\right\}-
	\mathbb{P}\left\{\Gamma^{\mbox{\tiny B},d}({\bf X}^{(d,\delta)})\neq Y\right\} &\leq& \sum_{k=1}^M \mathbb{E}\left[\Big|\phi_{d,k}({\bf X}^{(d,k)})-G_{m,d,k}({\bf X}^{(d,k)}) \Big| \bigg| \mathbb{D}_m\right].
	\end{eqnarray*}
\end{cor}
The proof of Corollary \ref{COR-1} is the same as that of Lemma \ref{LEM-3} and is obtained by conditioning on the training sample $\mathbb{D}_m$. 

\hfill $\Box$

\vspace{3mm}\noindent
The next lemma is a well-known result on the performance of the $L^1$-norm of kernel regression estimators.
\begin{lem}\label{LEM-4} {\it [Devroye et al (1996, Theorem 1); Gy\"{o}rfi et al (2002, Lemma 23.9).]}\\
	Let $(U, {\bf V})\in [-B, B]\times \mathbb{R}^d$, where $B<\infty$, and let $\phi({\bf v})=\mathbb{E}[U|{\bf V}={\bf v}]$ be the regression function. Let $\mathbb{D}_n=\{(U_1, {\bf V}_1),\dots,(U_n, {\bf V}_n)\}$ be the data (iid), where $(U_i, {\bf V}_i)  \stackrel{iid}{=}(U, {\bf V})$, and define
	$
		\widehat{\phi}_n({\bf v}) = \sum_{i=1}^n U_i \mathcal{K}(({\bf v}-{\bf V}_i)/{h_n})\,\big/\, \left\{n\, \mathbb{E}\left[\mathcal{K}({\bf v}-{\bf V})/h_n)\right]\right\},
	$
	where $\mathcal{K}: \mathbb{R}^d\to\mathbb{R}_+$ is regular. If $h_n\to 0$ and $nh_n^d\to \infty$, as $n\to\infty$, then for any distribution of $(U, {\bf V})$, any $\epsilon >0$, and $n$ large enough, 
	\[
	\mathbb{P}\left\{\mathbb{E}\left[\big|\widehat{\phi}_n({\bf V})- \phi({\bf V})\big|\,\Big|\, \mathbb{D}_n\right]> \epsilon \right\} \leq e^{-n\epsilon^2/(8B\rho)^2},
	\]
	where $\rho\equiv\rho(\mathcal{K})$ is a positive constant depending on the kernel  $\mathcal{K}$ only.
\end{lem}

\vspace{1mm}\noindent
PROOF OF THEOREM \ref{C}\\
Let $\Gamma^{\mbox{\tiny B},d}({\bf X}^{(d,\delta)})$ and $\Gamma^{\mbox{\tiny B}}({\bf X}^{(\delta)})$ be as in (\ref{E9B}) and (\ref{E9}), and observe that in view of part (iii) of Theorem \ref{B1} one has
\begin{eqnarray}
	&& \mathbb{P}\{\Gamma^{\mbox{\tiny B},d}({\bf X}^{(d,\delta)})\neq Y\} -\mathbb{P}\{\Gamma^{\mbox{\tiny B}}({\bf X}^{(\delta)})\neq Y\}\nonumber\\
	&& ~~ \leq~ \sum_{k=1}^M \mathbb{E} \bigg| \mathbb{E} \Big[(2Y-1) I\{\delta=k\}\Big|{\bf X}^{(k)}\Big]-  \mathbb{E} \Big[(2Y-1) I\{\delta=k\}\Big|{\bf X}^{(d,k)}\Big]\bigg|, \label{NEW12}
\end{eqnarray}
which follows upon taking $\varphi_k({\bf X}^{(k)})$, that appears in part (iii) of Theorem \ref{B1}, to be the same as the right side of (\ref{E8B}). Here, as before,  ${\bf X}^{(k)}=(X_1^{(k)}, X_2^{(k)},\, \dots)$ and ${\bf X}^{(d,k)}=(X_1^{(k)}, \dots, X_d^{(k)})$. Let $S_d^{(k)} = \mathbb{E} \big[(2Y-1) I\{\delta=k\}\big|{\bf X}^{(d,k)}\big]$ and
$S_{\infty}^{(k)} = \mathbb{E} \big[(2Y-1) I\{\delta=k\}\big|{\bf X}^{(k)}\big]$, and observe that for any $k=1,\dots,M$ and any $d_1<d_2$, one has $\mathbb{E}\big[S_{d_2}^{(k)}\big|X_1^{(k)}, \dots, X_{d_1}^{(k)}\big]$ $\stackrel{\mbox{\tiny a.s.}}{=} S_{d_1}^{(k)}$. Furthermore, $\sup_{d\geq 1}\big|S_d^{(k)}\big|\leq 1$. In other words, $\{S_d^{(k)}, d=1, 2, \dots\}$ is a martingale with respect to the increasing sequence of $\sigma$-fields, $\sigma(X_1^{(k)}, \dots, X_d^{(k)})$. Invoking the martingale convergence theorem (see, for example, Sec. 1.3 of Hall and Heyde (1980)), and arguing as in Biau et al. (2005), we find $S_d^{(k)}\to^{a.s.} S_{\infty}^{(k)}$, as $d\to\infty$. This fact together with the bound in (\ref{NEW12}) and an application of the dominated convergence theorem yield
$\mathbb{P}\{\Gamma^{\mbox{\tiny B},d}({\bf X}^{(d,\delta)})\neq Y\} -\mathbb{P}\{\Gamma^{\mbox{\tiny B}}({\bf X}^{(\delta)})\neq Y\} \to 0$, as $d\to\infty$. Consequently, for every $\epsilon>0$, and $n$ sufficiently large, there is a $d_{\epsilon}\in [1, d_n]$ such that $\mathbb{P}\{\Gamma^{\mbox{\tiny B},d}({\bf X}^{(d,\delta)})\neq Y\} -\mathbb{P}\{\Gamma^{\mbox{\tiny B}}({\bf X}^{(\delta)})\neq Y\} \leq \epsilon$ holds for all $d\geq d_{\epsilon}$ (recall $d_n\to \infty$ as $n\to \infty$). Therefore, for any $\widetilde{h}_k\equiv\widetilde{h}(n) \in \mathcal{H}_n$, $k=1,\dots,n$, satisfying the conditions of Theorem \ref{C}, any $\epsilon>0$, and $n$ large enough, one has
\begin{eqnarray}
&&
\mathbb{P}\left\{ \widehat{\Gamma}_n({\bf X}^{(\widehat{d},\delta)})\neq Y\,\Big|\,\mathbb{D}_n\right\} -
\mathbb{P}\left\{\Gamma^{\mbox{\tiny B}}({\bf X}^{(\delta)})\neq Y\right\} \nonumber\\
&&~~~~~~~~=~ \mathbb{P}\left\{ \widehat{\Gamma}_n({\bf X}^{(\widehat{d},\delta)})\neq Y\,\big|\,\mathbb{D}_n\right\} ~
- \inf_{1\leq d \leq d_n\,,\, h_1,\dots,h_M\in \mathcal{H}_n~} \mathcal{R}_m(d,h_1,\dots,h_M)\nonumber\\
&&~~~~~~~~~~~~~~+~ \inf_{1\leq d \leq d_n\,,\, h_1,\dots,h_M\in \mathcal{H}_n} \bigg\{\mathcal{R}_m(d,h_1,\dots,h_M) - \mathbb{P}\{\Gamma^{\mbox{\tiny B},d}({\bf X}^{(d,\delta)})\neq Y\} \nonumber\\
&&~~~~~~~~~~~~~~~~~~~~~~~~~~~~~~~~~~~~~~~~~~~~~~~+~ \mathbb{P}\{\Gamma^{\mbox{\tiny B},d}({\bf X}^{(d,\delta)})\neq Y\}\bigg\}-~ \mathbb{P}\left\{\Gamma^{\mbox{\tiny B}}({\bf X}^{(\delta)})\neq Y\right\}\nonumber\\
%&&~~~~~~~~\leq~ \mathbb{P}\left\{ \widehat{\Gamma}_n({\bf X}^{(\widehat{d},\delta)})\neq Y\,\big|\,\mathbb{D}_n\right\} ~- 
% \inf_{1\leq d \leq d_n\,,\, h_1,\dots,h_M\in \mathcal{H}_n~} \mathcal{R}_m(d,h_1,\dots,h_M)\nonumber\\
% &&~~~~~~~~~~~~+~ \inf_{h_1,\dots,h_M\in \mathcal{H}_n} \mathcal{R}_m(d_{\epsilon},h_1,\dots,h_M) - \mathbb{P}\{\Gamma^{\mbox{\tiny B},d_{\epsilon}}({\bf X}^{(d_{\epsilon},\delta)})\neq Y\} \nonumber\\
%&&~~~~~~~~~~~~~~~~~~~~~~~~~~~~~~~~~~~~~~~~~~~~~~~+~ \mathbb{P}\{\Gamma^{\mbox{\tiny B},d_{\epsilon}}({\bf X}^{(d_{\epsilon},\delta)})\neq Y\}-~ \mathbb{P}\left\{\Gamma^{\mbox{\tiny B}}({\bf X}^{(\delta)})\neq Y\right\}\nonumber\\
&&~~~~~~~~\leq~ \mathbb{P}\left\{ \widehat{\Gamma}_n({\bf X}^{(\widehat{d},\delta)})\neq Y\,\big|\,\mathbb{D}_n\right\} ~
- \inf_{1\leq d \leq d_n\,,\, h_1,\dots,h_M\in \mathcal{H}_n~} \mathcal{R}_m(d,h_1,\dots,h_M)\nonumber\\
&&~~~~~~~~~~~~~~+~ \mathcal{R}_m(d_{\epsilon},\widetilde{h}_1,\dots,\widetilde{h}_M)
- \mathbb{P}\{\Gamma^{\mbox{\tiny B},d_{\epsilon}}({\bf X}^{(d_{\epsilon},\delta)})\neq Y\} \nonumber\\
&&~~~~~~~~~~~~~~+~ \epsilon \label{AD1}
\end{eqnarray}
Now, in view of lemmas \ref{LEM-1} and \ref{LEM-2}, as $n\to\infty$, we have
\begin{equation}\label{AD3}
\mathbb{P}\left\{ \widehat{\Gamma}_n({\bf X}^{(\widehat{d},\delta)})\neq Y\,\big|\,\mathbb{D}_n\right\} ~
- \inf_{1\leq d \leq d_n\,,\, h_1,\dots,h_M\in \mathcal{H}_n~} \mathcal{R}_m(d,h_1,\dots,h_M)\longrightarrow^{\mbox{\small a.s.}} 0,
\end{equation}
Next, define
\[
\widetilde{\phi}_{m,d,h_k}({\bf x}) =
\frac{\widehat{\phi}_{m,d,h_k}({\bf x})}{m\cdot \mathbb{E}\left[\mathcal{K}_k\left(\frac{{\bf x}-{\bf X}^{(d,k)}}{h_k}\right)\right]}\,,
\]
where $\widehat{\phi}_{m,d,h_k}({\bf x})$ is as in (\ref{E10}), and observe that the classifier $\Gamma_m^{d}$ in (\ref{E11}) can alternatively be written as
$\Gamma_m^{d}({\bf X}^{(d,\delta)})=\sum_{k=1}^M I\{\delta=k\}I\big\{\widetilde{\phi}_{m,d,h_k}({\bf X}^{(d,k)})>0\big\}$. Therefore, by Corollary \ref{COR-1}, 
\begin{eqnarray}
\mathcal{R}_m(d_{\epsilon},\widetilde{h}_1,\dots,\widetilde{h}_M)
- \mathbb{P}\{\Gamma^{\mbox{\tiny B},d_{\epsilon}}({\bf X}^{(d_{\epsilon},\delta)})\neq Y\}  
&\leq& \sum_{i=1}^M \mathbb{E}\left[\Big|\phi_{d_{\epsilon},k}({\bf X}^{(d_{\epsilon},k)})-\widetilde{\phi}_{m,d_{\epsilon},\widetilde{h}_k}({\bf X}^{(d_{\epsilon},k)}) \Big| \bigg| \mathbb{D}_m\right] \nonumber\\
&&\longrightarrow^{\mbox{\small a.s.}} 0, ~~\mbox{as}~ n\to\infty,\label{AD4}\\
&&\mbox{(by Lemma \ref{LEM-4} and the Borel-Cantelli lemma)} \nonumber
\end{eqnarray}
where $\phi_{d_{\epsilon},k}$ is given by (\ref{E8B}). Therefore, in view of (\ref{AD1}), (\ref{AD3}), and (\ref{AD4}), for any $\epsilon>0$,
\begin{eqnarray*}
\lim_{n\to \infty} \left[\mathbb{P}\left\{ \widehat{\Gamma}_n({\bf X}^{(\widehat{d},\delta)})\neq Y\,\Big|\,\mathbb{D}_n\right\} -
\mathbb{P}\left\{\Gamma^{\mbox{\tiny B}}({\bf X}^{(\delta)})\neq Y\right\}\right] &\leq& \epsilon\,,
\end{eqnarray*}
almost surely. This completes the proof of Theorem \ref{C}. 

\hfill $\Box$\\

\vspace{0.5mm}\noindent
{\bf References}
\begin{description}
	\item[\rm Abraham, C.,] 
	Biau, G., Cadre, B., 2006. On the kernel rule for functional classification. AISM 58, 619-633.
	\item[\rm Abrams, D.,] 
	Goldma,n A., Launer, C., Korvick, J., Neaton J., et al. (1994). A comparitive trial of didanosine or zalcitabine after treatment with zidovudine in patients with human immunodeficiency virus infection. NEJM, 10, 657-662.
	\item[\rm Alonso, A,] 
	Casado, D., López-Pintado, S., Romo, J., (2014). Robust functional supervised classification for time series. J. Classification 31, 3, 325-350.
	\item[\rm Berlinet, A.,] 
	Biau, G., Rouviere, L., 2008. Functional classification with wavelets. Annales de l'Institut de statistique de l'universit\'e de Paris 52, 61-80.
	\item[\rm Biau, G.,] 
	Bunea, F., Wegkamp, M.H., 2005. Functional classification in hilbert spaces. IEEE T. Inform. Theory. 51, 2163-2172.
	%\item[\rm Bravo,\,F.,]
	%2015.\,Semiparametric estimation with missing covariates. J. Multivar.\,Anal.\,139, 329-346.
	\item[\rm Brooks-Gunn, J.,] 
	Duncan, G., Maritato, N. (1997). Poor families, poor outcomes: the well-being of children and youth. In: Duncan GJ, Brooks-Gunn J, editors. Consequences of Growing up Poor. Russell Sage Foundation; New York, pp. 1–17.
	\item[\rm Bugni,]
	F., 2012. Specification test for missing functional data. Economet. Theor. 28, 959-1002.
	\item[\rm Cai, T.,]
	Hall, P., 2006. Prediction in functional linear regression. Annals of Statistics, 34, 2159-2179.
	\item[\rm Case, A.,]
	Lubtosky, D., Paxson, C., 2002. Economic status and health in childhood: the origins of the gradient. Am. Econ. Rev. 92, 1308-1334.
	\item[\rm C\'erou, F.,] 
	Guyader, A., 2006. Nearest neighbor classification in infinite dimensions. ESAIM-Probab. Stat. 10,340-355.
	%\item[\rm Chenouri S,] 
	%Mojirsheibani M, Montazeri Z (2009) Empirical measures for incomplete data with applications. Electron J Stat 3:1021-1038.
	\item[\rm Chang, C.,] 
	Chen, Y., Ogden, R.T., 2014. Functional data classification: a wavelet approach. Computation. Stat. 29, 1497-1513.
	\item[\rm Cox, L.,] 
	Rouff, J., Svendsen, K., Markowitz, M., Abrams, D., CPCRA (1998). Community advisory boards:  their role in AIDS clinical trials. Health and Social Work, 4, 290-297.
	\item[\rm Cuevas, A.,] 
	Febrero, M., Fraiman, R., 2007. Robust estimation and classification for functional data via projection-based depth notions. Computation. Stat. 22, 481-496.
	\item[\rm Dai, X.,] 
	M{\"u}ller, H.-G., 2017. Optimal Bayes classifiers for functional data and density ratios. Biometrika 104, 3, 545-560.
	\item[\rm Delaigle, A.]
	Hall, P., 2013. Classification using censored functional data. J Am Stat Assoc 108, 1269-1283.
	\item[\rm Delaigle, A.]
	Hall, P., 2012. Achieving near perfect classification for functional data.  J Royal Stat Soc B 74, Part 2,  267-286.
	\item[\rm Delaigle, A.,]
	Hall, P., Bathia, N., 2012. Componentwise classification and clustering for functional data. Biometrika 99, 299-313.
	%\item[\rm Demirdjian, L.,] 
	%Mojirsheibani, M., 2017. Kernel classification with missing data and the choice of smoothing parameters. Stat. Pap. doi:10.1007/s00362-017-0883-y.
	\item[\rm Devroye, L.,]
	Gy\"{o}rfi, L., Lugosi, G., 1996. A Probabilistic Theory of Pattern Recognition. Springer, New York.
    \item[\rm Eun, H.,] and Choi, M. (2017). Relationship between structural changes of brain and cognitive function and survival rate in alzheimer's disease patients. Alzheimers Dement, 13, pp 1367. 
	\item[\rm Ferraty, F.,]
	Vieu, P., (2006). Nonparametric functional data analysis, Theory and practice. Springer, New York.
	\item[\rm Ferraty, F.,] 
	Vieu, P., 2003. Curves discrimination: a nonparametric functional approach. Comput. Stat. Data An. 4, 161-173.
	\item[\rm Gy\"{o}rfi, L.,]
	Kohler, M., Krzyzak, A., Walk, H., 2002. A distribution-free theory of non-parametric regression. Springer, New York.
	\item[\rm Hall, P.,]
	Heyde, C.C., 1980. Martingale limit theory and its application. Academic Press.
	\item[\rm Hall, P.,]
	Horowitz, J.L., 2007. Methodology and convergence rates for functional linear regression. Annals of Statistics, 35, 70-91.
	\item[\rm Hall, P.,] 
	Poskitt, D.S., Presnell, B., 2001. Functional data-analytic approach to signal discrimination. Technometrics 43, 1-9.
	\item[\rm He, W.,]  Goodkind, D., Kowal, P. U.S. Census Bureau, International Population Reports, P95/16-1, An Aging World: 2015, U.S. Government Publishing Office, Washington, DC, 2016.
	\item[\rm Kraus, D.,] 
	2015. Components and completion of partially observed functional data. J. R. Stat. Soc. B. 77, 777-801.
	\item[\rm Leng, X.,] 
	M{\"u}ller, H.-G., 2006. Classification using functional data analysis for temporal gene expression data. Bioinformatics 22, 68-76.
	\item[\rm L\'opez-Pintado,] 
	S., Romo, J., 2006. Depth-based classification for functional data, in DIMACS Ser. Discrete M. 72, 103-120.
	\item[\rm Meister, A.,]
	2016. Optimal classification and nonparametric regression for functional data. Bernoulli 22, 1729-1744.
	%\item[\rm Menga Y,]
	%Lianga J, Qian Y (2016) Comparison study of orthonormal representations of functional data in classification. Knowl-Based Syst 97:224-236.
	\item[\rm Mojirsheibani, M.,]
	Shaw, C., 2018. Classification with incomplete functional covariates. Stat. \& Probab. Lett. 139, 40-46.
	%\item[\rm Mojirsheibani, M.,] 
	%Montazeri, Z., 2007. Statistical classification with missing covariates. J. R. Stat. Soc. B. 69, 839-857.
	%\item[\rm Pawlak,]
	%M., 1993. Kernel classification rules from missing data. IEEE T. Inform. Theory 39,979-988.
	%\item[\rm Rachdi, M.,] Vieu, P., 2007. Nonparametric regression for functional data: Automatic smoothing parameter selection. J. Stat. Plan. Infer. 137, 2784-2801.
	%\item[\rm Reese, T.,] Mojirsheibani, M., 2017. On the Lp norms of kernel regression estimators for incomplete data with applications to classification. Stat. Method. Appl. 26, 81-112.
	\item[\rm Mosler, K.,] Mozharovskyi, P., 2017. Fast DD-classification of functional data. Stat. Papers 58, 1055-1089.
	\item[\rm Mueller, C.,] Stubbs, B., Banerji, S., Stewart, R., Perera, G. (2017). Concomitant use of anticholinergic medication attenuates benefits of cholinesterase inhibitors in alzheimer's disease:  a large cohort study of survival and cognitive function. Alzheimers Dement, 13, pp 851.
	%\item[\rm National] Institutes of Health (2018). HIV/AIDS:  The Basics.\\ 
	% URL https://aidsinfo.nih.gov/understanding-hiv-aids/fact-sheets/19/45/hiv-aids--the-basics.
	\item[\rm Rachdi, M.,] Vieu, P., 2007. Nonparametric regression for functional data: Automatic smoothing parameter selection. J. Stat. Plan. Infer. 137, 2784-2801.
	\item[\rm Sansone, G.,] 1969. Orthogonal Functions. Interscience, New York.
	\item[\rm Song, J.J.,] Deng, W., Lee, H.-J., Kwon, D., 2008. Optimal classification for time-course gene expression data using functional data analysis. Comput. Biol. Chem. 32, 426-432.
    \item[\rm van Buuren,] S., Groothuis-Oudshoorn, K. (2011). mice: Multivariate Imputation by
	Chained Equations in R. J. of Stat. Software, 45, 1-67.
	%\item[\rm van der Vaart,] A.W., Wellner, J.A., 1996. Weak Convergence and Empirical Processes with Applications to Statistics. Springer, New York.
	\item[\rm Yao, F.,]
	M{\"u}ller, H.-G., 2010. Functional quadratic regression. Biometrika, 97, 49-64.
    \item[\rm	Zygmund, A.,] 1959.
    Trigonometric Series I.  Cambridge Univ. Press.
\end{description}

\end{document}